\documentclass[a4paper,12pt]{article}
\usepackage{epsfig}
\usepackage{graphicx}
\usepackage{subfigure}
\begin{document}
\renewcommand{\baselinestretch}{01.35}
\renewcommand{\arraystretch}{0.666666666}
\parskip.2in
\newcommand{\hs}{\hspace{1mm}}
\newcommand{\nhat}{\mbox{\boldmath$\hat n$}}
\newcommand{\cmod}[1]{ \vert #1 \vert ^2 }
\newcommand{\mod}[1]{ \vert #1 \vert }
\newcommand{\pr}{\partial}
\newcommand{\fr}{\frac}
\newcommand{\ie}{{\em ie }}
\newcommand{\p}{\varphi}
\newcommand{\x}{\xi}
\newcommand{\xb}{\bar{\xi}}
\newcommand{\Vd}{V^{\dagger}}
\newcommand{\Ref}[1]{(\ref{#1})}
\title{\hbox{\hspace{18mm}{
\vbox{\Large{\bf Dynamical properties of a Soliton }
\break
\Large{\bf in a Potential Well}}
}}}

\author{
B. Piette\thanks{e-mail address: B.M.A.G.Piette@durham.ac.uk}
and W.J. Zakrzewski\thanks{e-mail address: W.J.Zakrzewski@durham.ac.uk}
\\ Department of Mathematical Sciences,University of Durham, \\
 Durham DH1 3LE, UK\\
}\date{October 2006}

\maketitle

\begin{abstract}
We analyse the scattering of a two-dimensional soliton on a potential well.
We show that this soliton can pass through the well, bounce back or become 
trapped and we study the dependence of the critical velocity 
on the width and the depth of the well. We also present a 
model based on a pseudo-geodesic approximation to the full system
which shows that the vibrational modes of the soliton play a crucial role
in the dynamical properties of its interactions with potential wells. 
\end{abstract}

\section{Introduction}

The scattering of a soliton with a well is an interesting phenomenon for 
several reasons. First of all, it is an example of the dynamical evolution of 
a soliton in an inhomogeneous medium, in which case the parameters of the 
model are 
functions of space. As is well known, when waves scatter on a well, they
can be partly reflected and partly transmitted. For solitons, the 
situation is more complicated as solitons cannot split and thus must 
either bounce, pass through or become trapped inside the well. This behaviour 
is very sensitive to the value of all the parameters of the model as well as 
to the initial conditions for the scattering.

The scattering of a soliton on a well is also interesting from a purely 
mathematical point of view. Very little is known about the general dynamical 
properties of non-integrable systems. Though many models have been studied 
numerically a general understanding of a  non-integrable dynamical system is 
still lacking. Several authors have claimed that the vibrational modes of 
solitons play an important role \cite{Kevrekidis}\cite{PW} and we shall 
show that the scattering of a soliton on a well is yet another example where
this is indeed the case.

The scattering of solitons on potential obstructions has also been studied 
in many other papers. In particular, we have found a paper by Fei et al 
\cite{Fei} who studied the scattering of one dimensional lattice Sine-Gordon 
kinks on a point defect. They found an interesting dependence on the 
initial kink velocity. Their work was taken further by Goodman and 
Haberman \cite{Goodman} who explained some of the results of 
\cite{Fei} by introducing a low dimensional dynamical model
involving the kink parameters together with a variable describing the 
amplitude of the standing wave localised around the defect. 
We have also looked at the scattering of Sine-Gordon kinks on potential 
obstructions (wells and barriers) and we are planning to report our results 
soon.

In a recent paper written in collaboration with Dr. Brand \cite{PZB}, we 
presented some results of our first study of the scattering 
of a two-dimensional soliton on a well of finite depth and width. 
We studied this scattering in a Baby Skyrme model whose solitons are 
often called skyrmions. There we 
showed that a soliton of this model can scatter through the well or become 
trapped, 
depending on its initial speed. In this paper, we analyse this 
scattering in more detail using numerical simulations and we also present a 
model based on a pseudo-geodesic approximation to the full system to explain 
our numerical results.

In the first section, we present the results of our numerical studies of the 
scattering of a soliton on a well, in which we varied both the width and 
the depth of the well.
 In the second section, we present two models based on pseudo-geodesic 
approximations to the full system and then use them to describe the 
scattering
of the soliton with the well. We show that the vibrations of the soliton 
play a crucial role in its dynamical properties.

\section{Skyrme Model with a Potential Well}

In this paper we present the results of our study of the so called New Baby 
Skyrme model (NBS)
in $2+1$ dimensions \cite{Weidig}, in which the mass coefficient depends on 
the space variable $x$ so that it describes
 a well of width $L$, centred around $x=0$ and
running parallel to the $y$ axis.
The field of the model takes values on a 2 dimensional sphere $S^2$ and so
can be described by a 3 component vector 
$\vec{\phi}= (\phi_0, \phi_1, \phi_2)$ of unit length : $|\vec{\phi}| = 1$.  
The Lagrangian of the model is given by
\begin{eqnarray}
L &=&  {1\over 2} \phi_{\mu}^2 
  - {\lambda\over 4}\left(\phi_{\mu}\times \phi_{\nu} \right)^2
  - {\lambda(1+\alpha)\over 2} (1-\phi_0^2)\nonumber \\
\label{lag}
\end{eqnarray}
where we have used the covariant notation 
$\phi_{\mu}^2 = \phi_{t}^2 -\phi_{x}^2 - \phi_{y}^2$ and where 
\begin{eqnarray}
\alpha &= a \qquad &-L/2 < x < L/2 \nonumber\\
\alpha &= 0 \qquad  &\hbox{elsewhere}
\end{eqnarray}
The appearance of $a \ne 0$ in a region of space corresponds to the 
introduction of a potential obstruction.
When $a > 0$ we have a potential barrier and the parameter $a$ describes its 
height;  when $a$ is negative, we have a well and $|a|$ 
describes the depth of the well. Notice that $a \ge -1$. 
Given the interesting results in the well case that were reported in 
\cite{PZB}, in this paper we consider the case $a<0$.

Outside the well, $\alpha=0$, and so the parameters of the Skyrme and 
potential terms
are identical. In this parametrisation, also used in \cite{PW}, the 
parameter $\lambda$ is dimensionless and hence the soliton size varies
with $\lambda$ only through its shape.
When the parameters of the two terms in the Lagrangian differ, as is inside 
the well, one
can rescale the space coordinates $x$, $y$ and the model parameters to
make them equal.
Calling $\lambda_1=\lambda$, the coefficient of the Skyrme term, and
$\lambda_2 = \lambda\,(1+a)$, the coefficient of the potential term, we
perform the scaling
\begin{eqnarray}
x_i &=& \sigma X_i\qquad i=1,2,x=x_1,\quad y=x_2, \quad \nonumber\\
\tilde{\lambda}_1 &=& \frac{1}{\sigma^2} \lambda_1,\nonumber\\
\tilde{\lambda}_2 &=& \sigma^2 \lambda_2
\end{eqnarray}
and we see that if we choose $\sigma = (\lambda_1/\lambda_2)^{1/4}$
then $\tilde{\lambda}_1=\tilde{\lambda}_2=(\lambda_1\lambda_2)^{1/2}$ are 
the parameters of the model in dimensionless units.

Inside the well, $a<0$, and we have 
$\lambda_1 =\lambda$, $\lambda_2 =\lambda(1+a)$;
so after the scaling we thus have
\begin{eqnarray}
\lambda_{ef} &=& \lambda (1+\alpha)^{1/2}.\nonumber\\
x_{ef} &=& x (1+\alpha)^{1/4}.
\label{lambda}
\end{eqnarray}
We thus see that the effective value of the parameter, $\lambda_{eff}$, is
smaller inside the well. From this we can conclude that the energy of a static
soliton inside the well is smaller than that of a soliton outside it. 
Moreover, the soliton is broader inside the well, as seen in Figure 1 where we 
have presented the plots of the energy profiles of the soliton for various 
well depths.

\begin{figure}[htbp]
\unitlength1cm \hfil
\begin{picture}(8,8)
 \epsfxsize=8cm \put(0,0){\epsffile{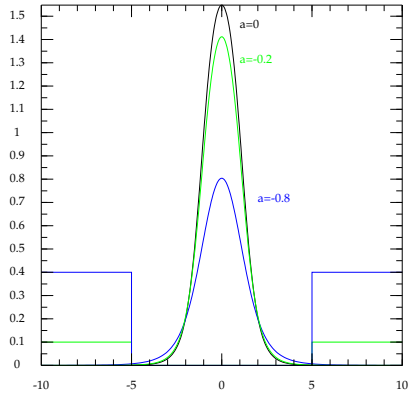}}
\end{picture}
\caption{\label{SolProf} Energy profile of a soliton in the well for wells 
of width $L=10$ and depth $a=0$, $a=-0.2$ $a=-0.8$. $\lambda=0.5$.
}
\end{figure}

\begin{figure}[htbp]
\unitlength1cm \hfil
\begin{picture}(8,8)
 \epsfxsize=8cm \put(0,0){\epsffile{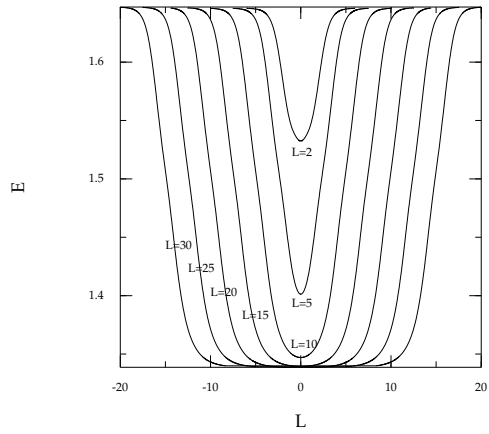}}
\end{picture}
\caption{\label{potEn_L10} Potential energy of the soliton as a function of 
its position when $\lambda=0.5$. The well, of depth $a=-0.8$, is located 
between $x=-L/2$ and $x=L/2$. 
}
\end{figure}

As the energy of a soliton is smaller inside the well, we see that
the well attracts the soliton. To understand the scattering of the soliton 
on such a well, it is useful to have an idea of the energy of the soliton as a 
function of its position relative to the centre of the well. Such an energy
profile is presented in figure \ref{potEn_L10} for various values of $L$ when 
$a=-0.8$. These energy profiles were obtained by solving numerically the
time evolution of the soliton as its slides down into the well having added 
a friction term to the equations of motion to slow the soliton and absorb its 
kinetic energy.
We see from figure \ref{potEn_L10} that due to the spatial extension of the 
soliton, the well is effectively smoothed out.


\subsection{Numerical Solutions}
As the Lagrangian (\ref{lag}) is invariant under Lorentz boosts along the $y$ 
direction we can, without any loss of generality, restrict ourselves to 
studying head on collisions of the soliton with the well. Moreover, the 
relative orientation of the soliton, a phase, does not play any role 
as the well does not depend on that phase.

\begin{figure}[htbp]
\unitlength1cm \hfil
\begin{picture}(16,16)
 \epsfxsize=8cm \put(0,8){\epsffile{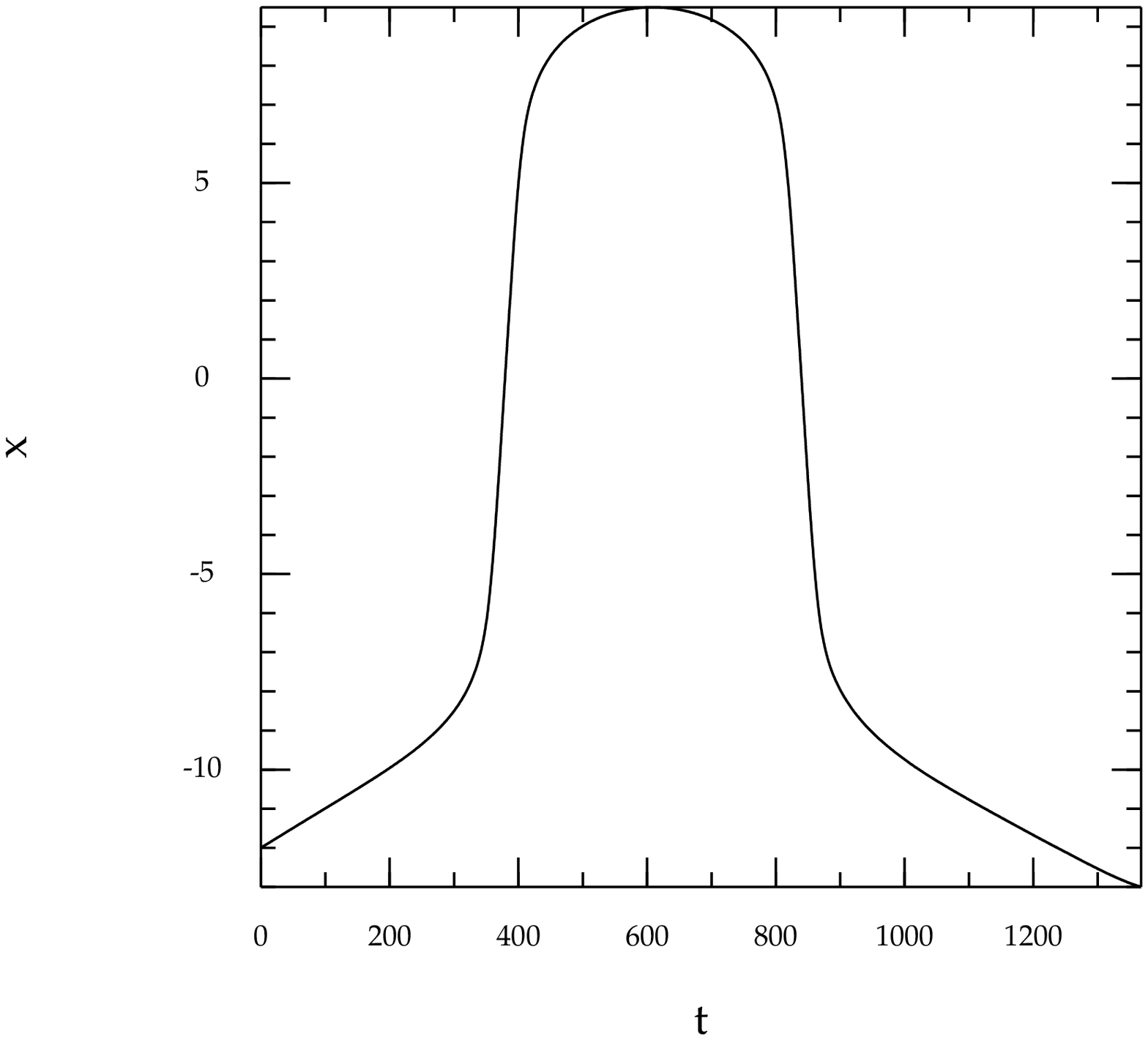}}
 \epsfxsize=8cm \put(8,8){\epsffile{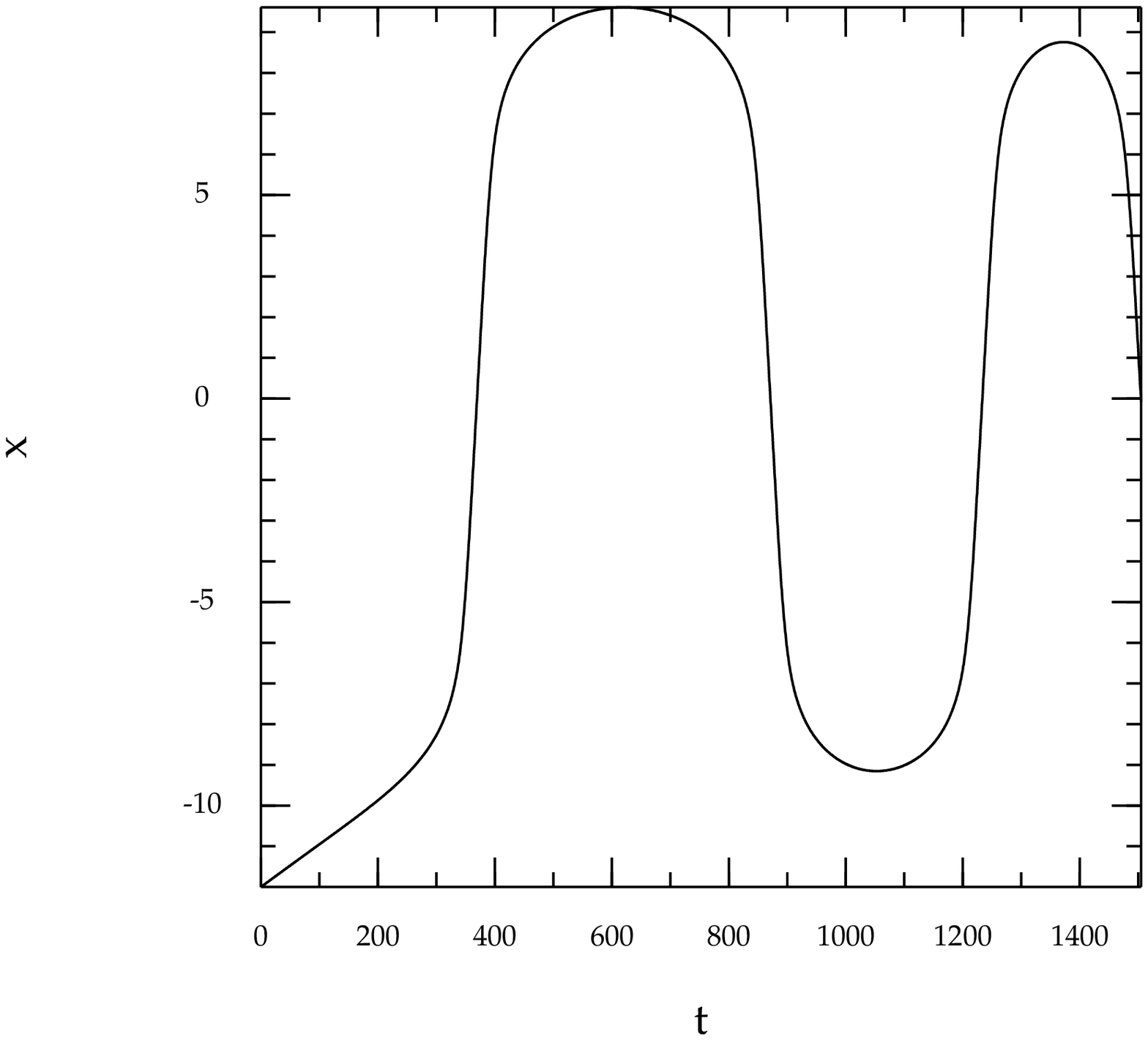}}
 \epsfxsize=8cm \put(0,0){\epsffile{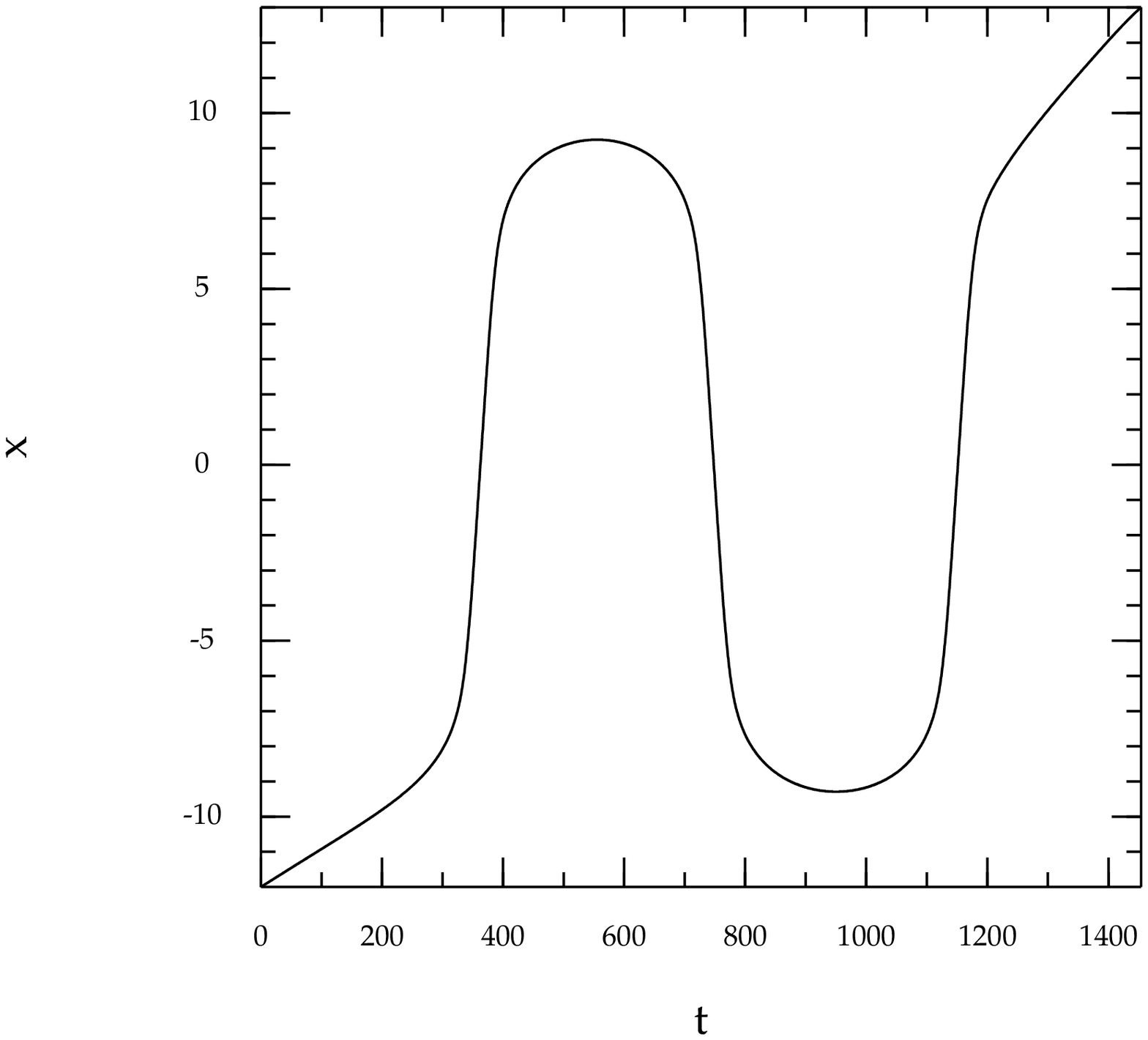}}
 \epsfxsize=8cm \put(8,0){\epsffile{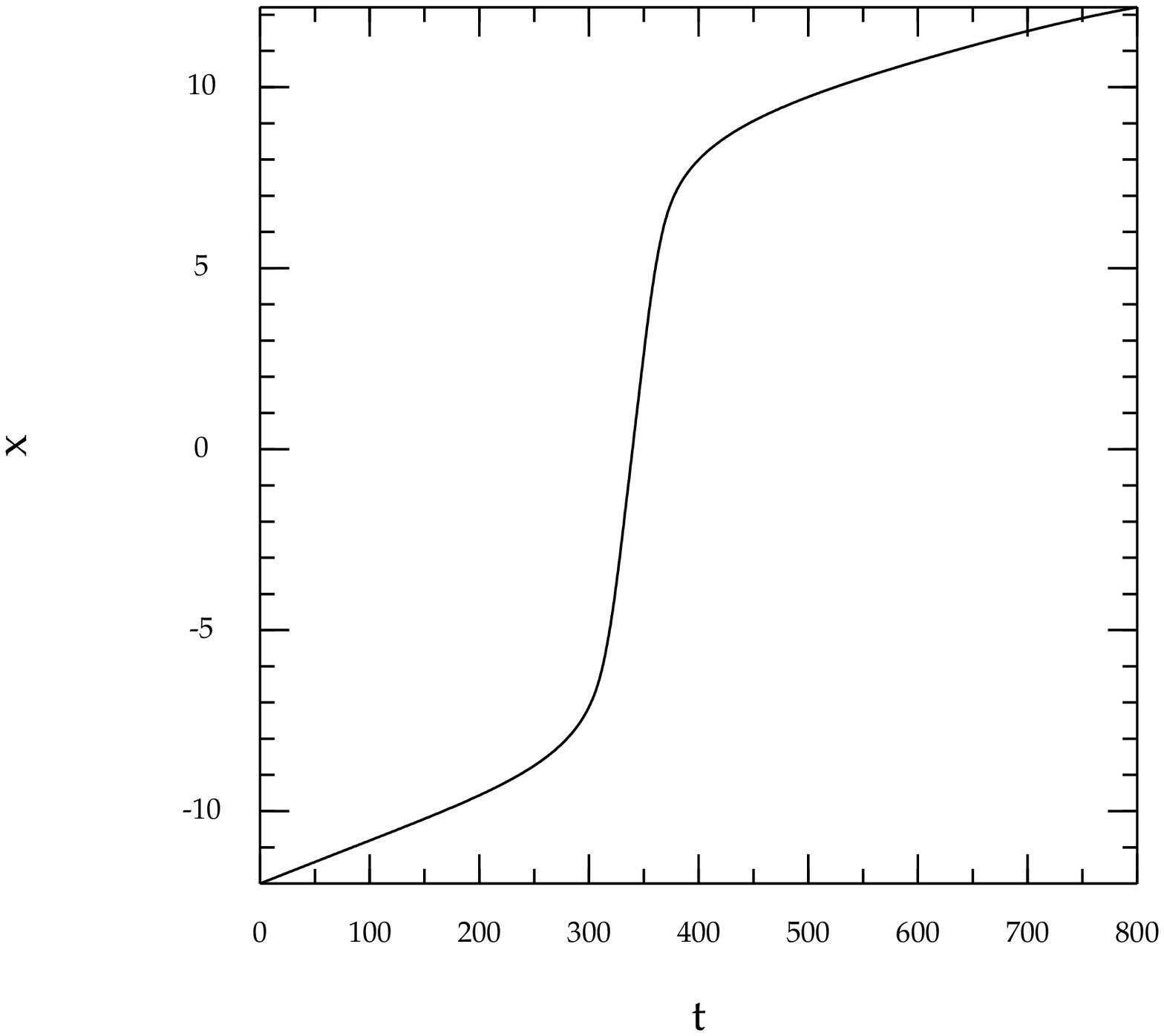}}
\put(4,8){a}
\put(12,8){b}
\put(4,0){c}
\put(12,0){d}
\end{picture}
\caption{\label{traj_L10} Trajectory of a soliton during the scattering for 
a well of width $L=10$ and depth $a=-0.2$ for $\lambda=0.5$ and a) $v=0.0102$, 
b) $v=0.0106$, c) $v=0.0109$, d) $v=0.012$.
}
\end{figure}

As reported in \cite{PZB}, the behaviour of the soliton when it scatters on 
the well depends very much on its initial speed. At very low speeds, the
soliton falls into the well and gets trapped. For larger speeds, on the other 
hand, it passes through the well and emerges nearly unaltered
(Fig. \ref{traj_L10} d). 
For intermediate values of the speed, the trajectory of the soliton is 
extremely sensitive to its initial speed. In some instances, the soliton 
falls into the well, bounces against the opposite wall of the well, moves 
backwards and then escapes from the well in the direction it came from 
(Fig. \ref{traj_L10} a). 
For other values of the speed, the soliton bounces twice inside the well before
escaping from the well (Fig. \ref{traj_L10} c). The ranges of initial 
speeds at which these 
different scatterings occur are very narrow and often are, maybe always,
separated by regions where the soliton is trapped inside the well
(Fig. \ref{traj_L10} b). We suspect that the soliton can sometimes bounce 
more than twice inside the well, but due to the time required to perform the 
numerical simulations and the narrowness of the region where this could occur,
we have not tried to find such a scattering.

As the soliton falls inside the well, it readjusts its shape and as a result 
it becomes excited {\it i.e.} it starts vibrating. 
The soliton of the NBS model has several vibrational modes with frequencies 
which depend on the parameter $\lambda$ of the model\cite{PW}.
When the frequencies of oscillation of these vibrational modes are above the 
mass threshold, the vibrational modes are coupled to the radiation modes of 
the model resulting in a damping of the oscillations through radiation.
When the frequencies are below the mass threshold, the soliton vibrations are 
stable, at least, in the linear approximation, when the amplitudes are 
sufficiently small. 

In the NBS model there are two stable vibrational modes when $\lambda<1.1$. 
They correspond to the shape mode describing radial contraction and expansion, 
and the scattering mode involving alternating elongation and contraction along 
orthogonal 
directions. The frequencies of these two modes are very close to each other 
and they cross the mass threshold \cite{PW} around $\lambda\approx 0.27$. 

Regardless of the value of $\lambda$, the soliton vibrations are always
excited when it falls into the well and so some of the potential energy is 
always converted into the kinetic energy of these vibrations. 
This explains why, at small speeds, the soliton 
gets trapped inside the well. For large velocities, the kinetic energy of the
translation is always large enough for the soliton to climb out of the well, 
and so the soliton just sails through it. In between these two extremes, the 
energy is transferred between the different degrees of freedom resulting in
a complex dynamical system which we have to understand if we want 
to explain what we have seen in this process.

Very qualitatively, one can see that to escape from the well, the phases of
the oscillations of the solitons must be such that the potential energy of the
system is large enough just when the soliton gets to the edge of the well, 
while trying to climb out of it. 

\begin{figure}[htbp]
\unitlength1cm \hfil
\begin{picture}(16,8)
 \epsfxsize=8cm \put(0,0){\epsffile{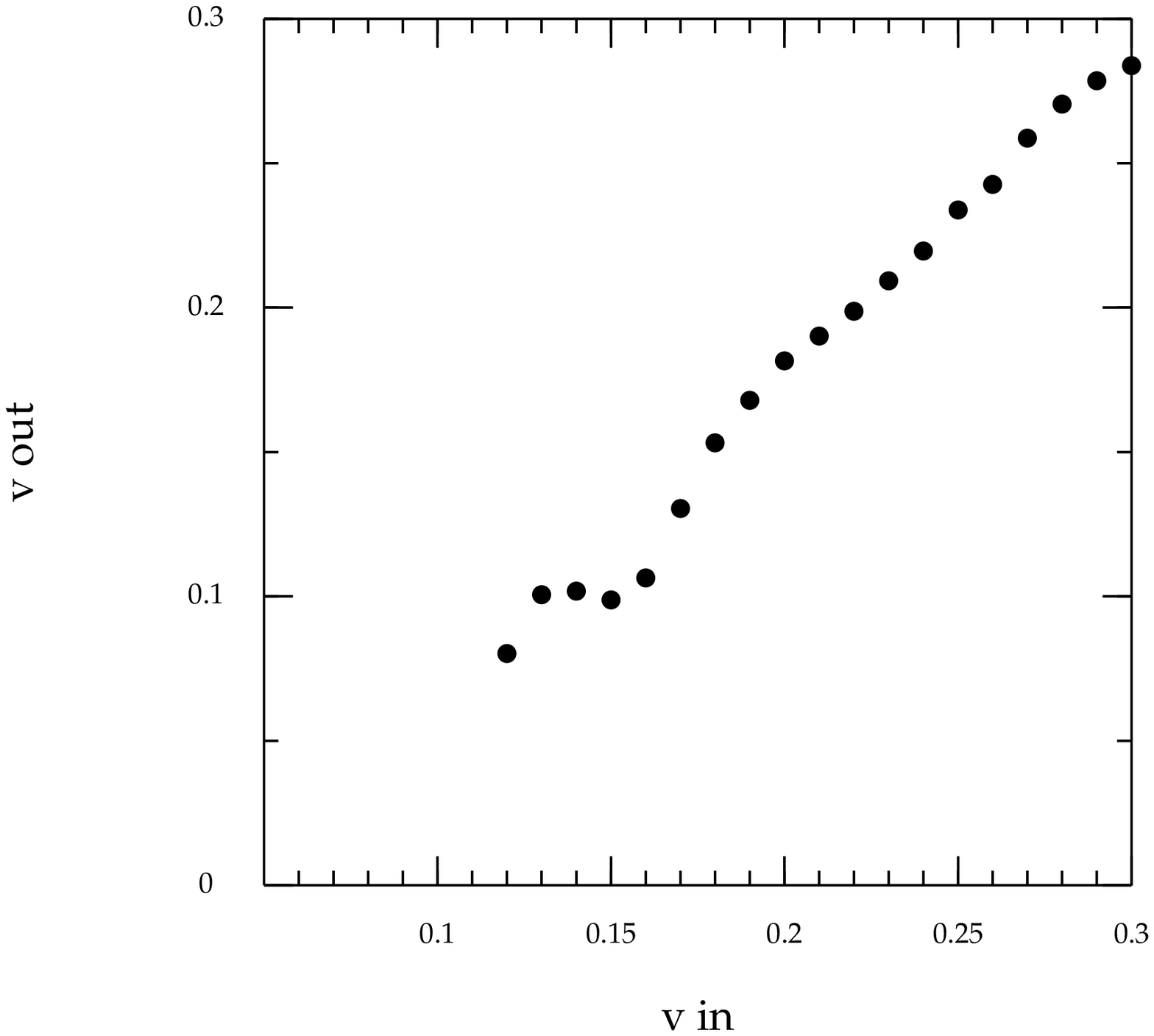}}
 \epsfxsize=8cm \put(8,0){\epsffile{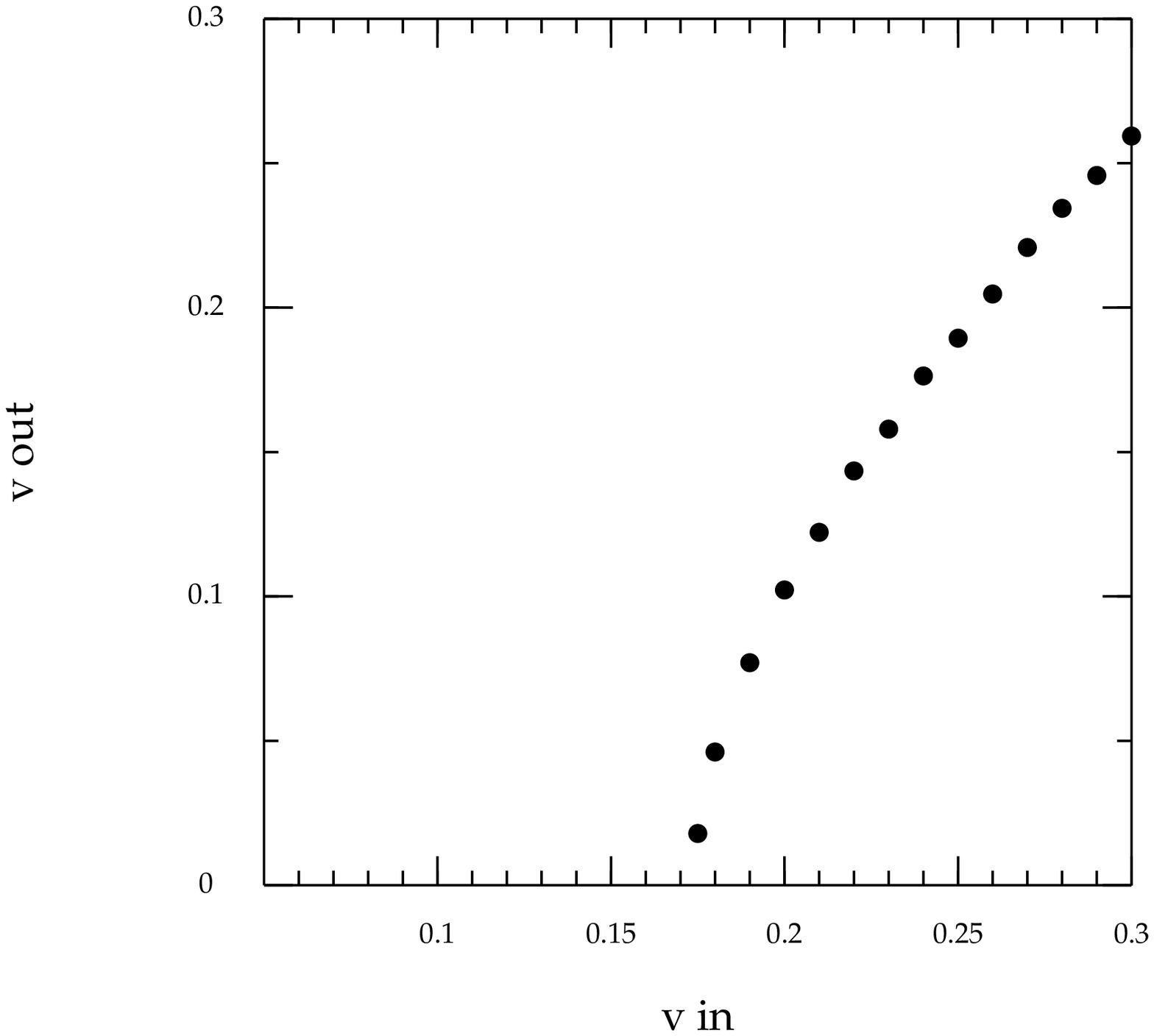}}
\put(4,0){a}
\put(12,0){b}
\end{picture}
\caption{\label{vout_L10k} Speed of the soliton after scattering on a well
of width $L=10$ and depth $a=-0.8$ for a) $\lambda=0.2$. b) $\lambda=0.5$.
}
\end{figure}

In figure \ref{vout_L10k} we present the outgoing speed of the soliton as a 
function of the incoming speed for two values of the parameter $\lambda$. One 
clearly sees that there is indeed a critical velocity below which the soliton 
can be trapped by the well. Moreover the outgoing 
speed tends to increase with the incoming speed except for small 
fluctuations around $\lambda=0.2$.

\begin{figure}[htbp]
\unitlength1cm \hfil
\begin{picture}(16,8)
 \epsfxsize=8cm \put(0,0){\epsffile{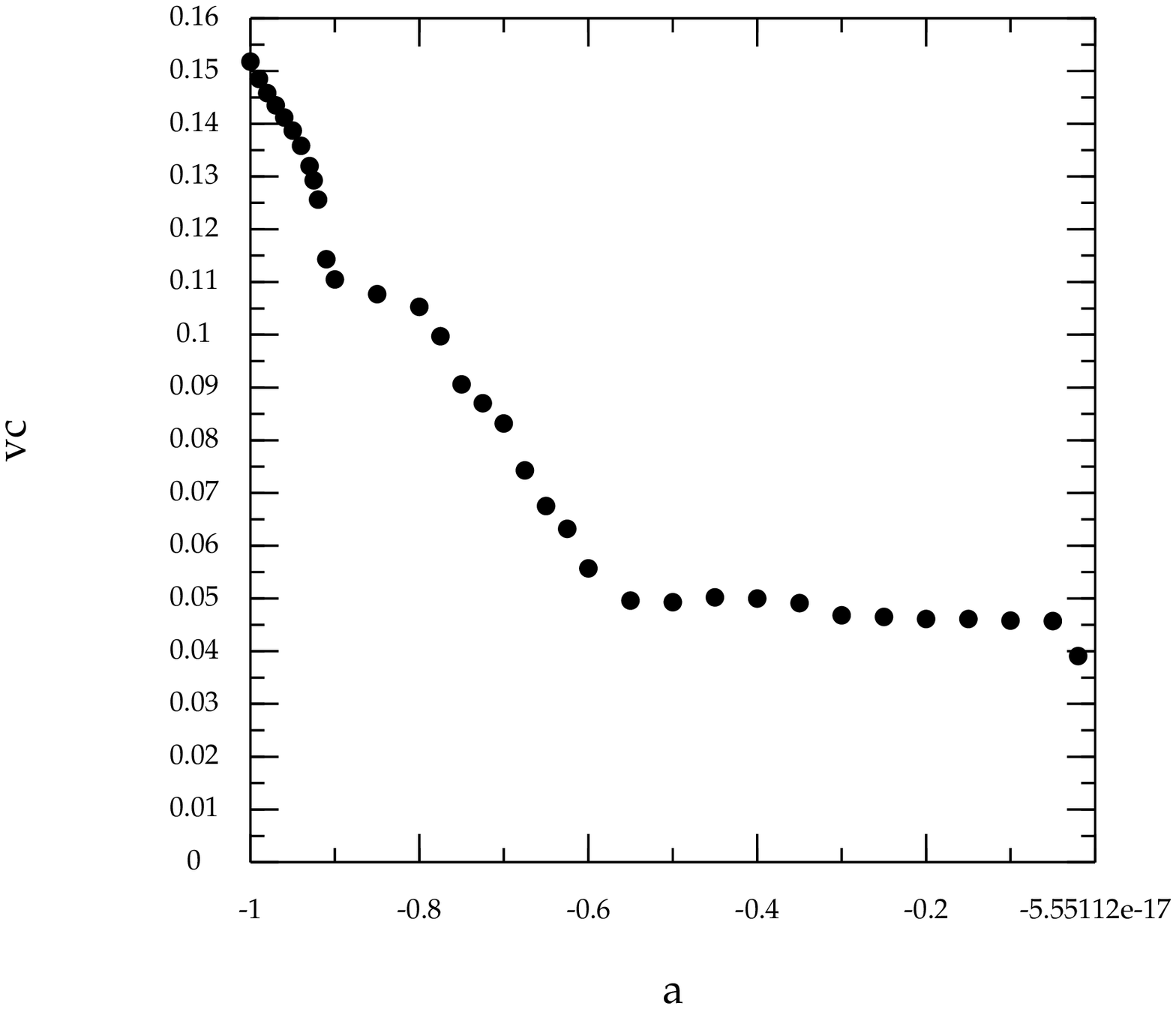}}
 \epsfxsize=8cm \put(8,0){\epsffile{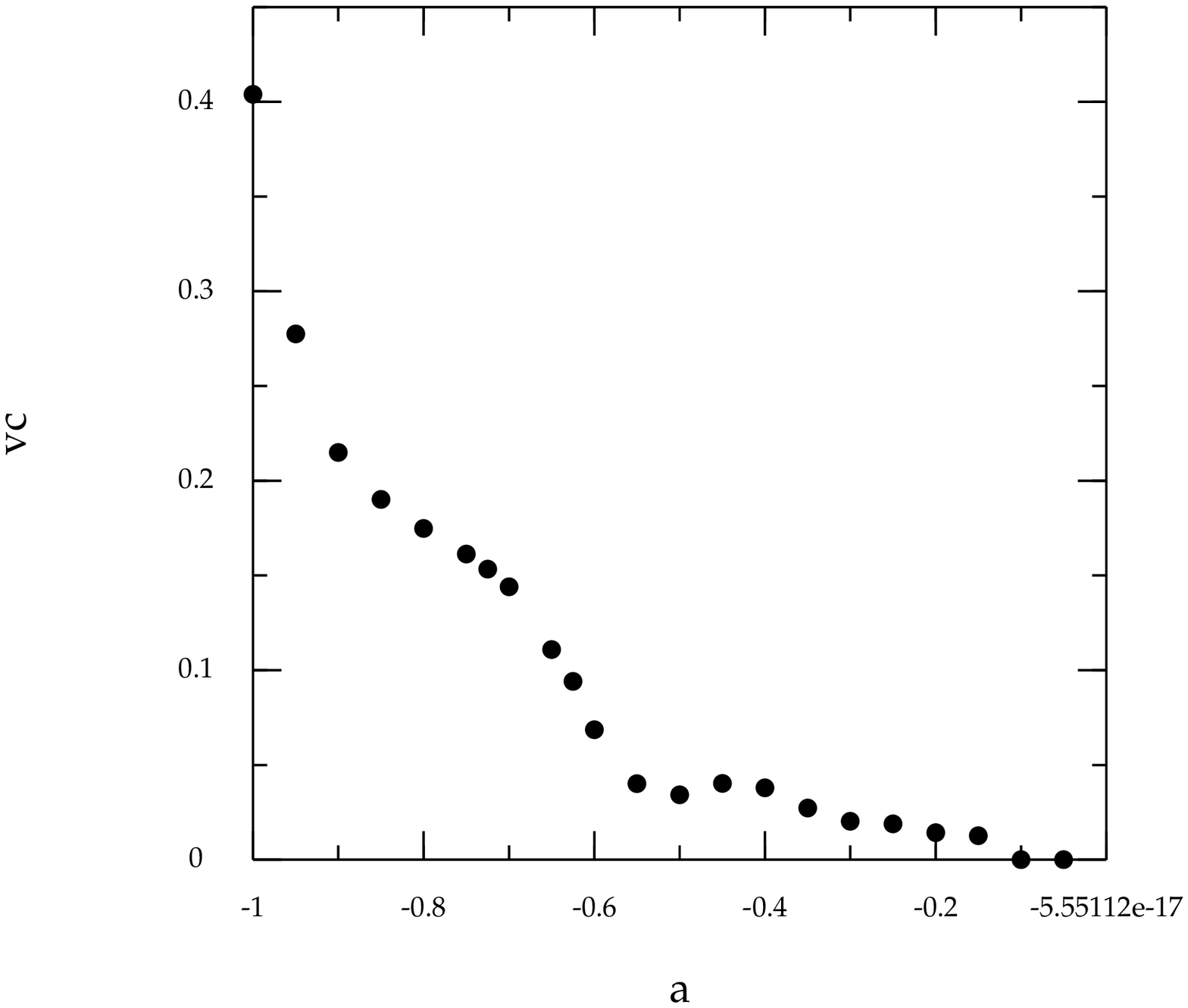}}
\put(4,0){a}
\put(12,0){b}
\end{picture}
\caption{\label{vc_L10k} Critical velocity as a function of 
the depth of the well $a$ for $L=10$ and a) $\lambda=0.2$. b) $\lambda=0.5$.
}
\end{figure}

\begin{figure}[htbp]
\unitlength1cm \hfil
\begin{picture}(8,8)
 \epsfxsize=8cm \put(0,0){\epsffile{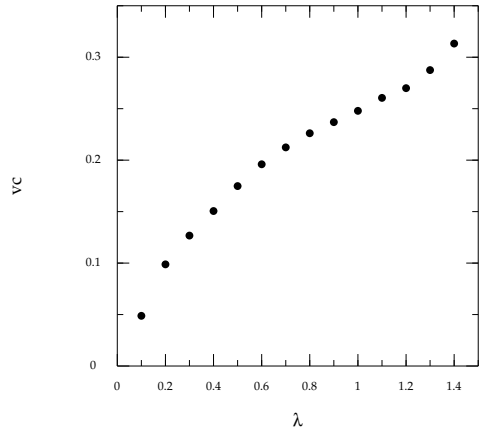}}
\end{picture}
\caption{\label{vc_L10a} Critical velocity as a function of 
the parameter $\lambda$ for $L=10$ and $a=-0.8$.
}
\end{figure}

\begin{figure}[htbp]
\unitlength1cm \hfil
\begin{picture}(16,8)
 \epsfxsize=8cm \put(0,0){\epsffile{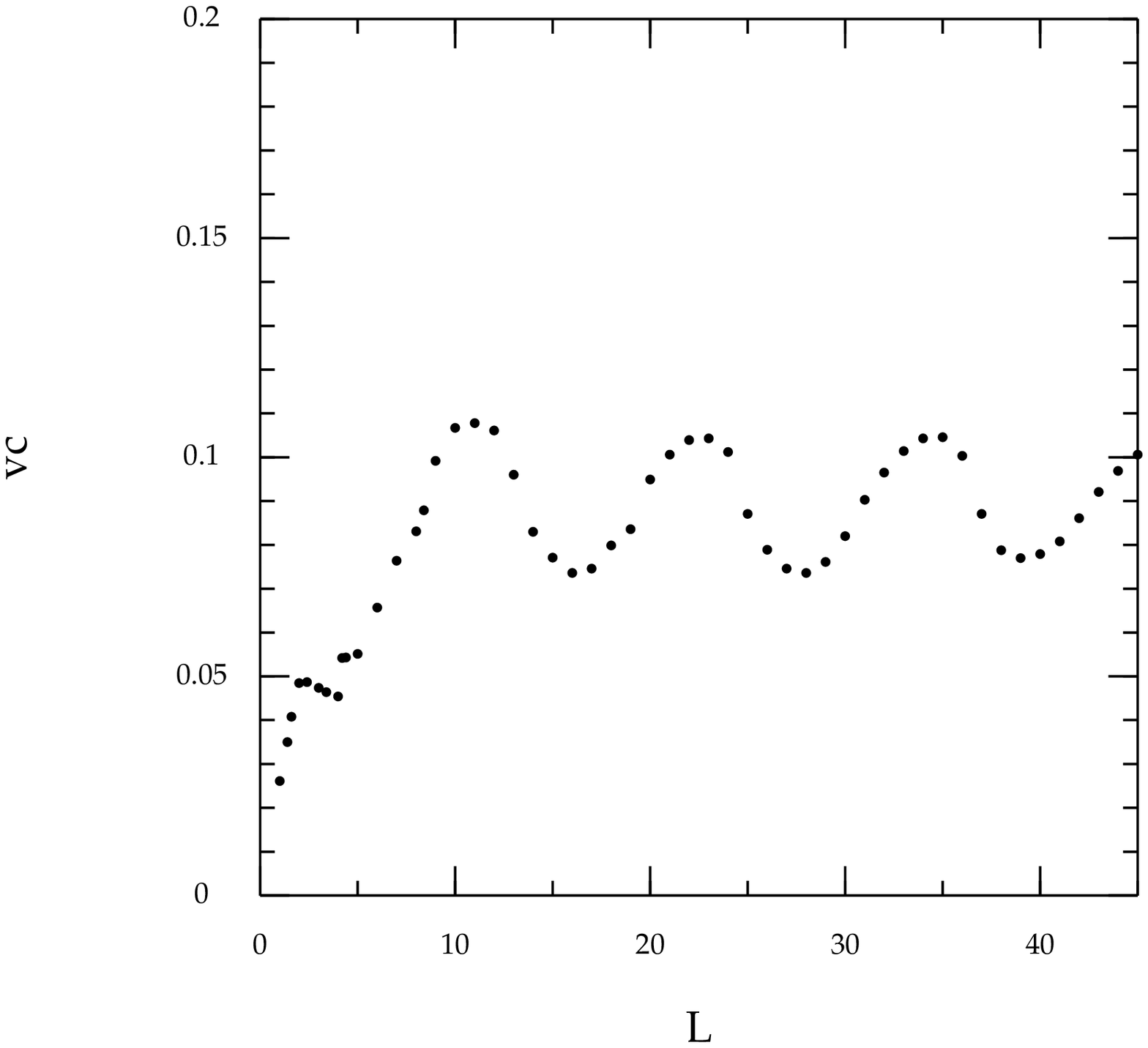}}
 \epsfxsize=8cm \put(8,0){\epsffile{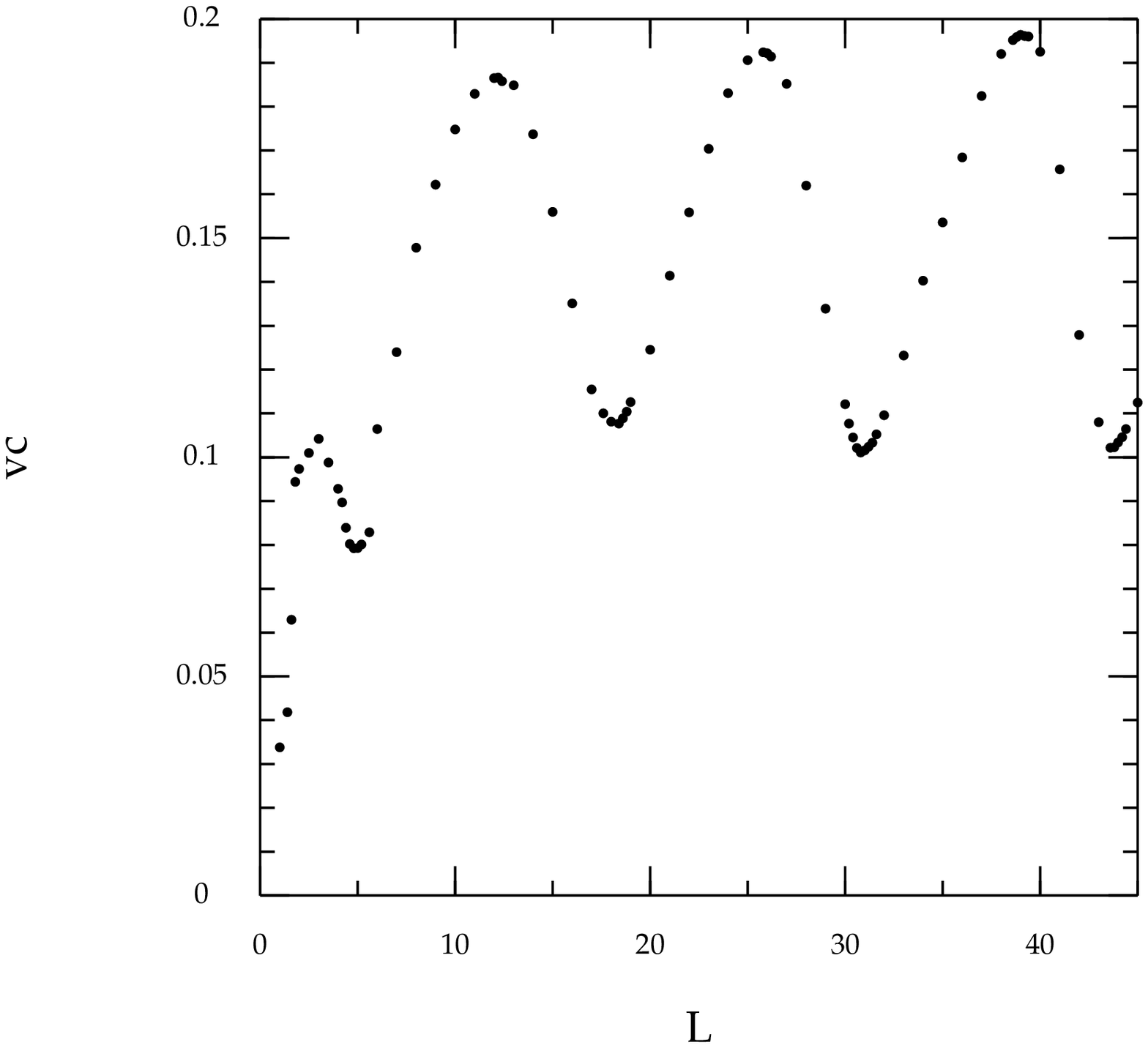}}
\put(4,0){a}
\put(12,0){b}
\end{picture}
\caption{\label{vc_a-0.8k} Critical velocity as a function of 
the well width $L$ for $a=-0.8$ and a) $\lambda=0.2$. b) $\lambda=0.5$.
}
\end{figure}

To analyse the scattering of the soliton on the well, we have computed the 
critical velocity of the soliton for different values of the parameters 
$\lambda$ and for different wells. In what follows, we define the critical 
speed
as the speed above which the soliton is never trapped by the well.
The results are shown in Figures \ref{vc_L10k}-\ref{vc_a-0.8k}.

From figure \ref{vc_L10k} we see that, as expected, the critical velocity
increases when the well becomes deeper. The curve is relatively linear, but 
it exhibits a few humps. In the case $\lambda=0.2$ we also observe a plateau 
where $v_c \approx 0.05$ in the region $-0.5 < a < 0.05$. We believe this 
comes from the fact that every soliton vibrational mode radiates in this 
model and that even 
for a shallow well, the soliton looses some energy when it falls into 
the well,  thus explaining the relatively large value of its critical speed.

In figure \ref{vc_L10a} we present the critical velocity as a function of the 
parameter $\lambda$ and we see that, at least for our choice of
the well, the critical velocity increases with $\lambda$.

In figure \ref{vc_a-0.8k} we present the dependence of the critical velocity
on the width of the well $L$. When the well is narrower than the soliton 
itself, 
$L <10$, the critical velocity increases with $L$ except for a small hump in 
the region $3 < L <5$.
When $L > 10$ the curve exhibits oscillations with very regular amplitudes and 
periods which depend on $\lambda$. 
These oscillations can be explained by the fact that to climb out of the 
well, the soliton must be in the correct phase of vibration. 
The phase is itself dictated by the frequency of vibration
and the time needed to cross the well which is directly related to 
the speed of the soliton. So as the length of the well changes, the
speed of the soliton must also change to allow it to escape, but if the length
of the well is extended by the distance travelled by the soliton during one 
period of oscillation, then the critical speed returns to its previous value.

\begin{figure}[htbp]
\unitlength1cm \hfil
\begin{picture}(8,8)
 \epsfxsize=8cm \put(0,0){\epsffile{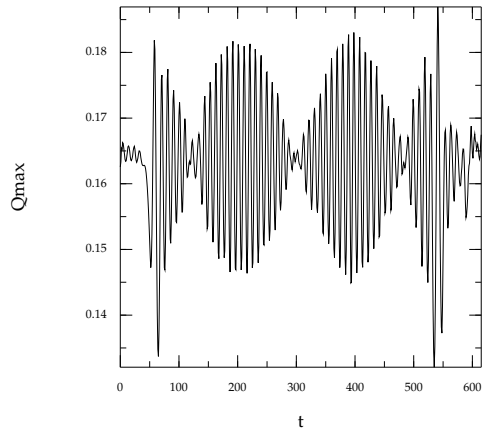}}
\end{picture}
\caption{\label{QOsc} Oscillation of the topological charge density maximum
during a one bounce scattering for $\lambda=0.5$, $L=2$, $a=-0.8$ and 
$v=0.0970313$.
}
\end{figure}

In Figures \ref{vc_a-0.8k} a and b, the periods in $L$ are, approximately, 
$12$ and $13$, respectively. The average speeds are, respectively, $v=0.09$ 
and 
$v=0.15$ and the periods of oscillation are thus given by $T=133$ and
$T=87$, respectively. These periods are about 6 times larger than the 
frequency of the oscillation of the soliton \cite{PW}. So even though our 
explanation looks reasonable, the actual data do not support it however,
looking at figure \ref{QOsc} we note that the oscillations of the maximum of
the topological charge density
show that several modes of oscillations are excited. Our original explanation 
has just been  
too naive and we need to take into account the role of several vibrational 
modes.
Incidentally, Figure \ref{QOsc}
corresponds to the case where the soliton bounces in the well once, around 
$t=300$ and then escapes from the well, around t=$600$, going back to where it 
came from.

In the next section we present a model, based on 
a pseudo-geodesic approximation.
This model shows that, while our initial picture is essentially correct, 
some details of it need a modification. This is due to the fact that
several degrees of freedom do indeed come into play in this process and so 
lead to a larger oscillation in $L$. 
This leads us to an improved pseudo-geodesic model which, 
qualitatively, correctly describes the scattering of the soliton with the well.

\subsection{Models based on a Pseudo Geodesic Approximation}
To explain the characteristics of the scattering of the NBS soliton with the 
well we present a model which is a generalisation of the geodesic 
approximation to the full process. 

Note that have we had an analytical expression to describe the evolution of 
the soliton as it interacts with the well we would have approximated the 
soliton by a rigid object evolving in an effective potential induced by 
the well. To describe the soliton 
vibrations, we would have then to consider a small perturbation around this 
``gradient flow'' and expand the Hamiltonian to quadratic order in this 
perturbation. This would have
led us to a dynamical system with zero modes and vibrational modes identical 
to those of the solitons, at least in the 
small amplitude limit, which would evolve in an effective potential describing 
the changes experienced by the soliton as it falls into the well. 

As we do not have an analytical expression for the time evolution of the 
soliton, we must thus adopt a 
different approach, and so we have decided to generalise the pseudo-geodesic 
approximation introduced 
in \cite{PW}, {\it i.e.} by trying to construct a dynamical system that has 
similar vibrational modes as the soliton and that evolves in a potential 
approximating 
the one experienced by the soliton falling into the well. 

As the soliton falls into the well, effectively, it goes down a 
potential more or less described by  Fig \ref{potEn_L10}. To model the 
vibrations
of the soliton, we have first tried to use a system made out of two masses 
separated by a finite distance, related to the size of the soliton, and linked 
together by a spring in such a way that the frequency 
of the normal mode of the system matches the main vibrational mode of the NBS 
soliton. 
Moreover, the distance between the masses and the spring tension have to be a 
function of the position of the centre of mass to model the parameter changes 
that occur as the soliton moves inside the well. 

To model the transition of the soliton between the inside and outside of the 
well we introduce the profile
\begin{eqnarray}
P(s;x) &=& \tanh(s\,(x-L/2)) + \tanh(s\,(-x-L/2)).
\label{profile}
\end{eqnarray}
where $s$ is a scale parameter which has to be fitted.

The potential in which the 2 masses evolve is then given by
\begin{eqnarray}
V(x) &=& E_0 + \frac{dE}{2}\, c\, P(s;x)\\
&& c = \left\{ \begin{array}{ll}
                   (L/10)^{1/4} & L<10\\
                              & \\
                   0          & L\ge 10
               \end{array}
       \right. \nonumber
\label{effPot}
\end{eqnarray}
where $E_0$ is the energy of the soliton outside the well and $E-dE$ is
the energy of the soliton when it is at rest inside the well. 
The parameter $c$ is a factor that is required to correct the potential depth 
when the well is too narrow, {\it i.e.} when only a part of the soliton is 
located inside the well.
The potential (\ref{effPot}) constitutes a good approximation to the curves 
in figure \ref{SolProf} when one takes $s\approx 0.5$. 
  
The equilibrium distance between the two masses should be a function of the 
position of the centre of mass, $x_{cm}=\frac{1}{2}(x_1+x_2)$, and so we 
set it to twice the radius of the soliton.
Defining $D_0(\lambda)$ as the equilibrium distance outside the well,
we choose the position dependant equilibrium distance as
\begin{equation}
D(x_{cm}) = D_0 - \frac{1}{2}d\, P(s_D;x_{cm}),
\label{Dx}
\end{equation}
where $s_{D}$ is a scale parameter and $d$ is the difference between the
equilibrium distance between the 2 masses inside and outside the well.

The Hamiltonian of our effective model is thus given by
\begin{equation}
H = \frac{1}{2} (M \dot{x_1}^2 + M \dot{x_2}^2) + \frac{k}{2}(x_2-x_1-D(x_{cm}))^2 
     +\frac{1}{2}(V(x_1)+V(x_2))
\end{equation}
where $2M=E_0$ is the energy of the soliton at rest and 
$k = M [0.3+0.63(1-\exp(-1.58 \lambda))]^{1/2}$ (cfr \cite{PW}) sets 
the vibration frequency of the soliton. 

In Table 1 we give some of 
the numerical values that we have used for this pseudo-geodesic approximation.

\begin{center}
\begin{table}[h]
\begin{tabular}{|l|l|l|l|l|}
\hline
                        & $E_0$   & $dE$      & $D_0$    & $d$\\
\hline
$\lambda=0.2$, $a=-0.8$ & 1.30701 & 0.149235  & 3.1611  & 1.73382\\
$\lambda=0.2$, $a=-0.2$ & 1.30701 & 0.0267838 & 3.1611  & 0.19758 \\
$\lambda=0.5$, $a=-0.8$ & 1.64923 & 0.312953  & 3.05734 & 1.64716\\ 
$\lambda=0.5$, $a=-0.2$ & 1.64923 & 0.0572152 & 3.05734 & 0.18602 \\ 
\hline
\end{tabular}
\caption{Parameter values for the 2 mass pseudo-geodesic model}
\end{table}
\end{center}

The equation for $x_1$ and $x_2$ are then given by
\begin{eqnarray}
M \ddot{x_1} &=& - k(x_2-x_1-D(x_{cm}))(-1 + \frac{1}{4}\,d\,G(s_{D};x_{cm}))
                -  \frac{dE}{4} G(s;x_1)
\nonumber\\ 
M \ddot{x_2} &=& - k(x_2-x_1-D(x_{cm}))(1 + \frac{1}{4} \,d\,G(s_{D};x_{cm}))
             -  \frac{dE}{4} G(s;x_2)
\end{eqnarray}

where
\begin{equation}
 G(s;x) = {\partial P(s;x) \over \partial x} 
        = s(\tanh^2(s\,(-x-L/2)) -\tanh^2(s\,(x-L/2))).
\end{equation}
 
We can also add a friction term proportional to $\dot{x_1}-\dot{x_2}$ to model 
the radiation of the soliton. In the new baby Skyrme model, the vibrations
of a single soliton are coupled to radiation when 
$\lambda < \lambda_c = 0.27$.
If we define 
\begin{equation}
\lambda_{eff} = \lambda -\frac{1}{2}  \lambda ((1+a)^{1/2}-1) P(s_{D};x)
\label{keff}
\end{equation}
we can then add the following friction force when $k_{eff} < k_c$:
\begin{eqnarray}
Fr_{x_1} &=&  -K_{fr} (\lambda_c - \lambda_{eff})(\dot{x_1}-\dot{x_2})
\nonumber\\ 
Fr_{x_2} &=&  -K_{fr} (\lambda_c - \lambda_{eff})(\dot{x_2}-\dot{x_1})
\end{eqnarray}
where $K_{fr}$ is a friction coefficient.

\begin{figure}[htbp]
\unitlength1cm \hfil
\begin{picture}(16,8)
 \epsfxsize=8cm \put(0,0){\epsffile{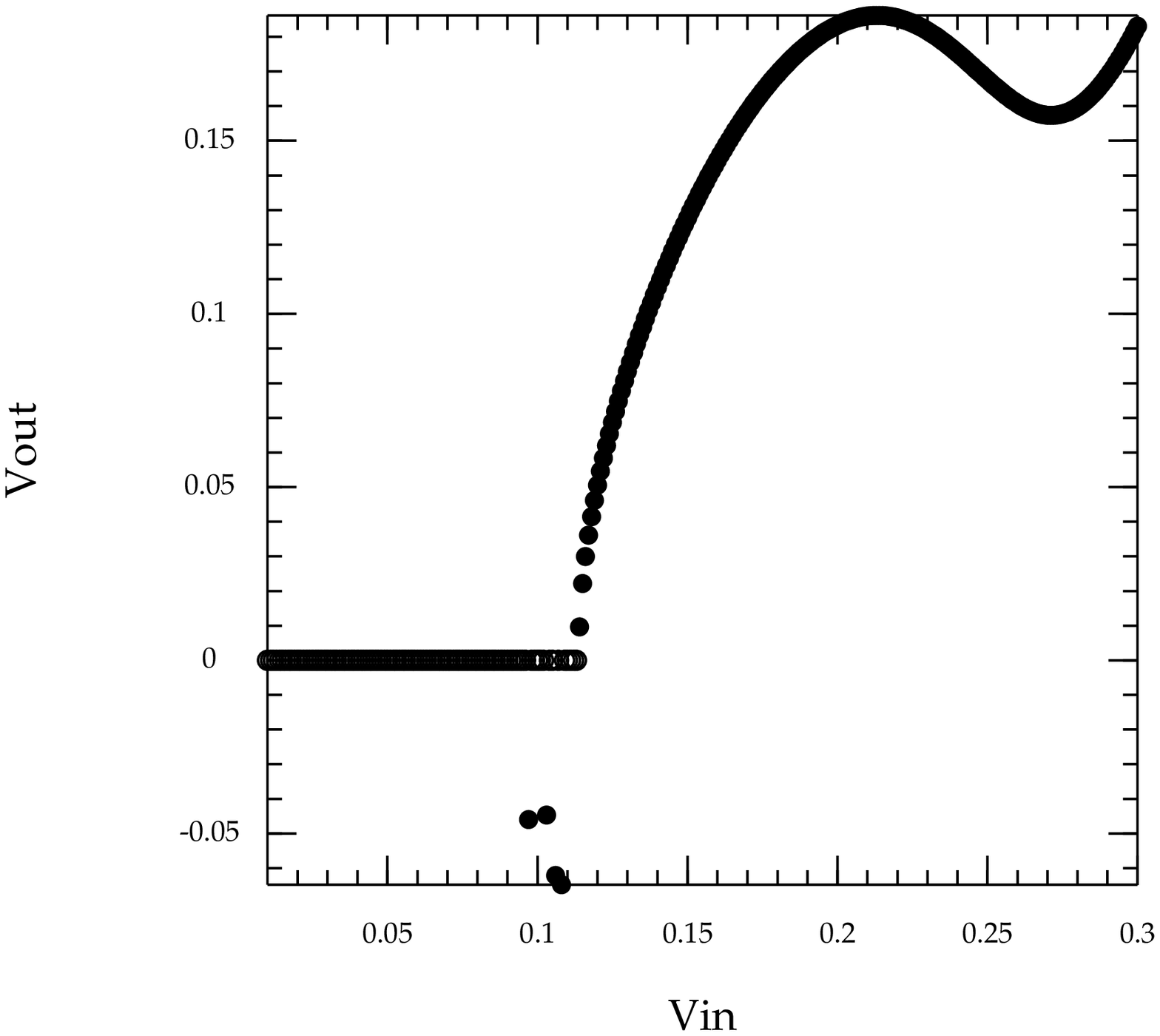}}
 \epsfxsize=8cm \put(8,0){\epsffile{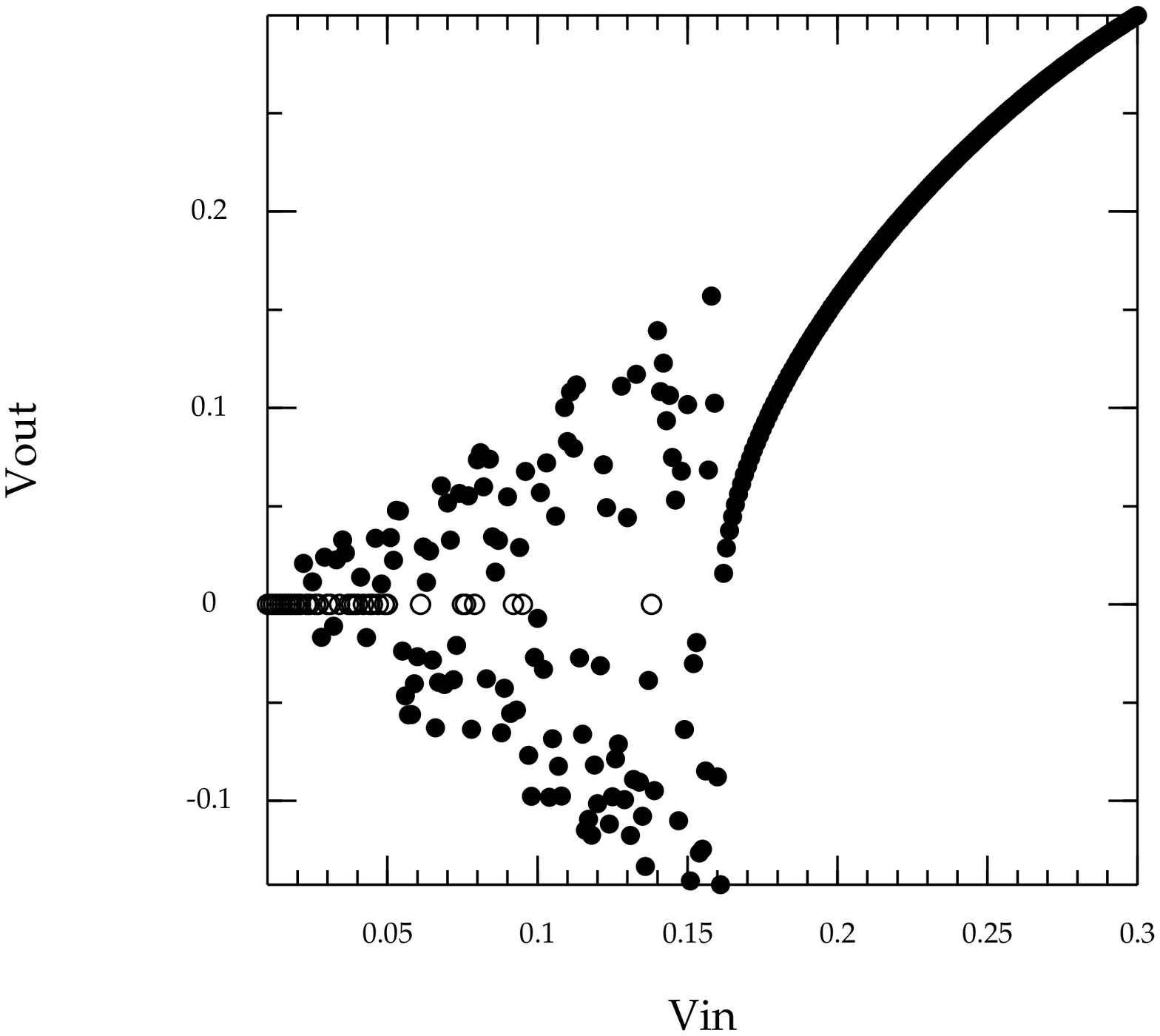}}
\put(4,0){a}
\put(12,0){b}
\end{picture}
\caption{\label{m2vout_L10k} Speed of a soliton in the 2 mass model after 
scattering on a well of width $L=10$ and depth $a=-0.8$ for 
a) $\lambda=0.2$, $s=s_D=1.6, K_{fr}=0.01$ 
b) $\lambda=0.5$, $s=0.5,s_D=1.6, K_{fr}=0$.
}
\end{figure}

In figure \ref{m2vout_L10k} we present the outgoing velocity of the soliton as 
a function of the incoming speed obtained in the model based on the 2 mass 
pseudo-geodesic approximation for the same cases as those of figure 
\ref{vout_L10k}. We note that when the speed is small enough the soliton is 
trapped. Sometimes, however, the soliton bounces several times in the well and 
eventually escapes from the well even when the initial speed is below the 
critical velocity. This is particularly clear from fig \ref{m2vout_L10k}.b 
where we see that the soliton can escape in  any direction and with almost 
any speed that is restricted by the conservation of energy.

\begin{figure}[htbp]
\unitlength1cm \hfil
\begin{picture}(16,8)
 \epsfxsize=8cm \put(0,0){\epsffile{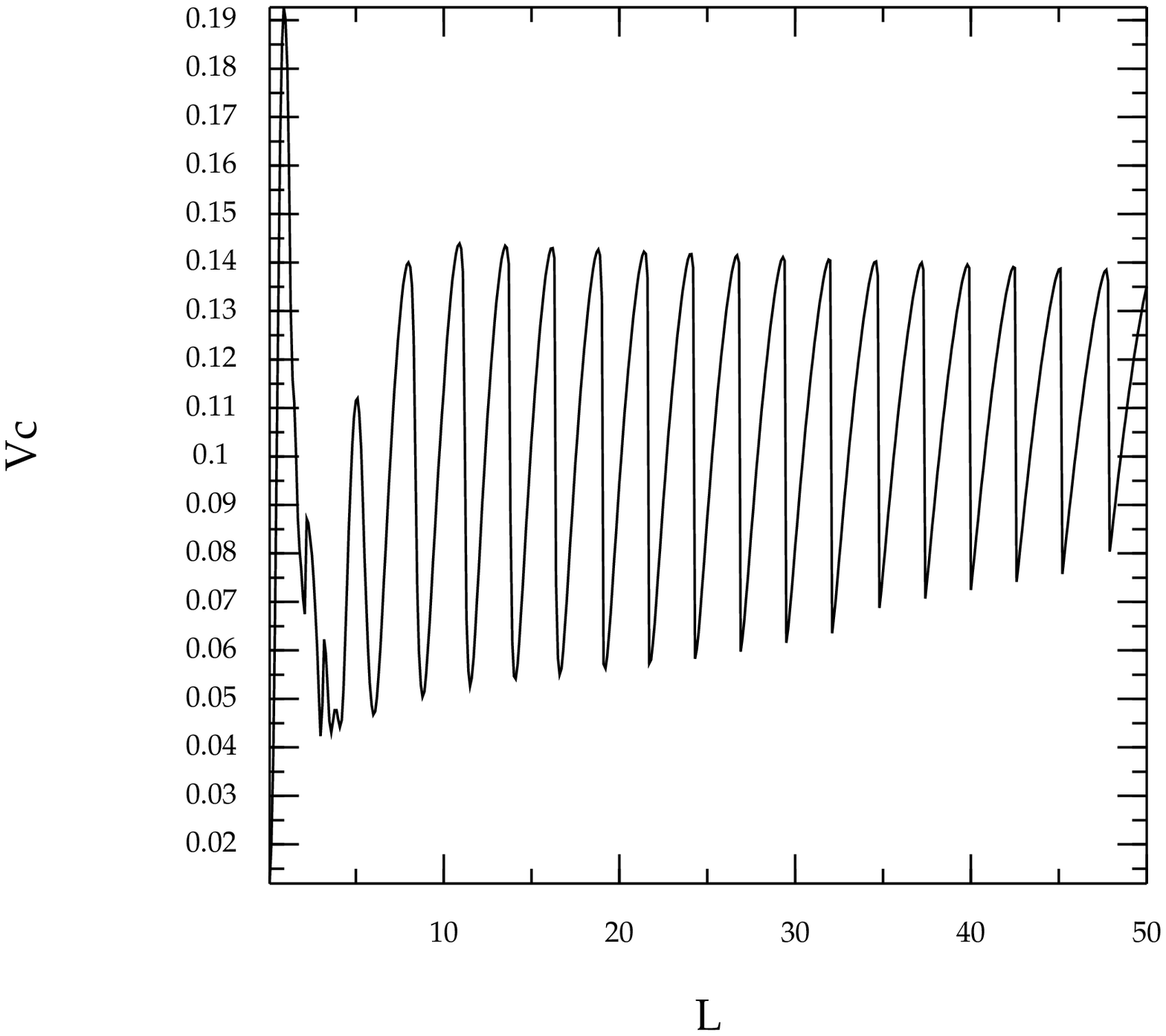}}
 \epsfxsize=8cm \put(8,0){\epsffile{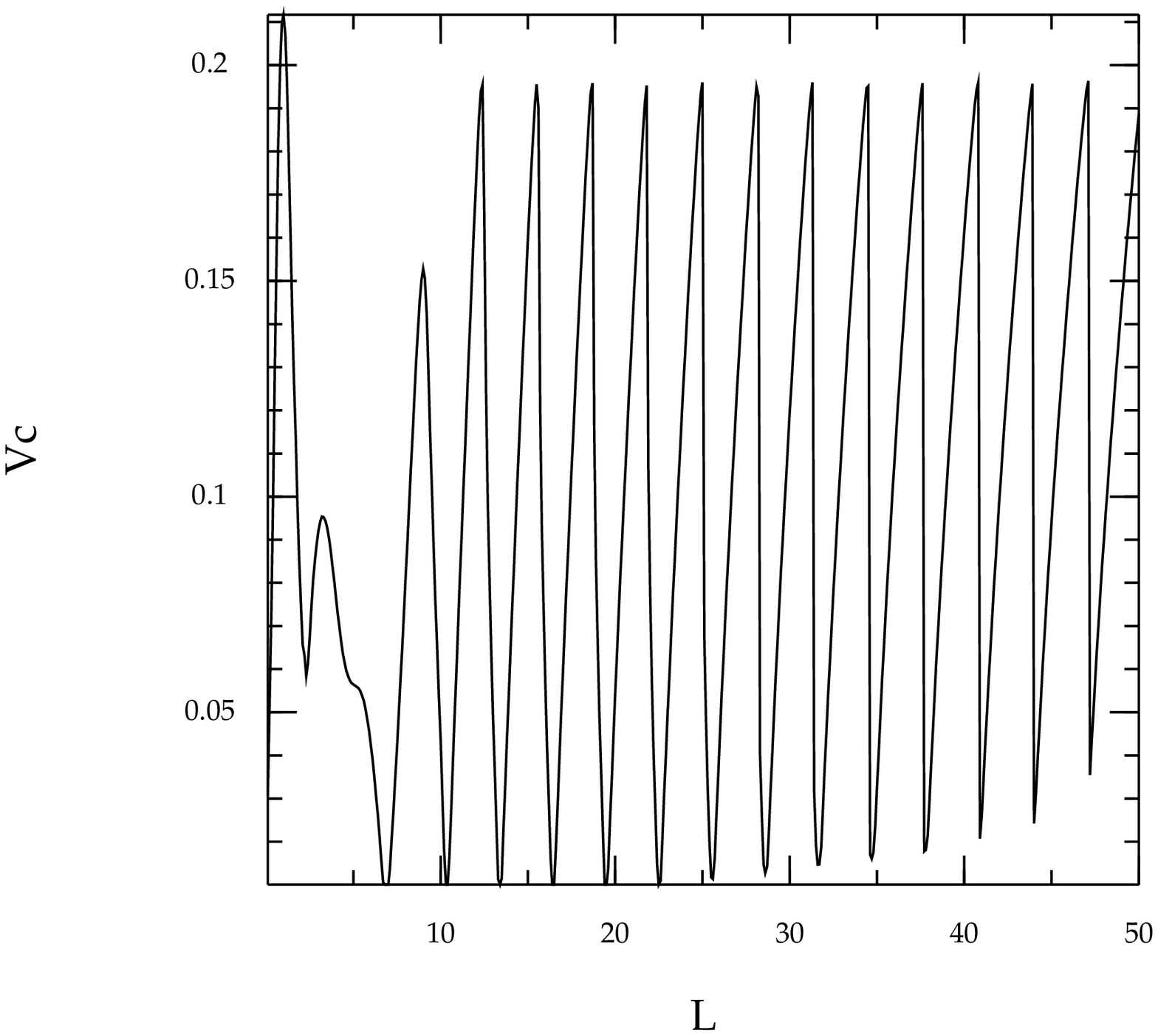}}
\put(4,0){a}
\put(12,0){b}
\end{picture}
\caption{\label{m2vc_a-0.8k} Critical velocity in the 2 mass model 
as a function of the well width $L$ for $a=-0.8$ and 
a) $\lambda=0.2$, $s=s_D=1.6, K_{fr}=0.01$ 
b) $\lambda=0.5$, $s=1.05,s_D=1.05, K_{fr}=0$.
}
\end{figure}

In figure \ref{m2vc_a-0.8k} we plot the critical velocity as 
a function of the well width $L$. We observe oscillations like those in figure 
\ref{vc_a-0.8k}, with a similar amplitude, but the period of these 
oscillations is about 
5 times too small. The oscillations can be explained by the fact
that to escape from the well, the soliton must be in the correct phase. This 
phase depends on the time spent by the soliton in the well, which itself 
depends on the 
width of the well. If the critical velocity is $0.1$, and as the frequency of 
the oscillation is roughly $20$ one would expect the period of oscillation to 
be around $2$ in our dimensionless units, and this is what we see in figure  
\ref{m2vc_a-0.8k}. The soliton on the other hand oscillates with a period 
roughly $5$ times larger (Fig. \ref{vc_a-0.8k}). This suggests that
for the soliton, several excited oscillation modes contribute to the 
build up of the correct phase. 

The NBS model, in fact, has several vibrational modes \cite{PW} and it is the 
superposition of all these modes that determines the phase of the total system.
In a complex system like the soliton, the extrema of the critical velocity 
as a function of $L$ will be determined by how close or how far the system
is from the ideal phase for achieving the escape from the well. The 
periodicity of $V_c(L)$ can thus differ from the exact periodicity of the 
full system, but nevertheless, the periods of oscillation of a system with 
several vibrational modes are usually larger than the lowest frequency, 
explaining why the oscillations in Fig \ref{vc_a-0.8k} 
are larger than the frequency of the lowest vibrational mode, 
which our previous naive argument expected to match  
the period of $V_c(L)$.

To show that this is indeed the case we next consider a model based on a 
multi-mode pseudo-geodesic approximation.

\begin{figure}[htbp]
\unitlength1cm \hfil
\begin{picture}(16,8)
 \epsfxsize=8cm \put(0,0){\epsffile{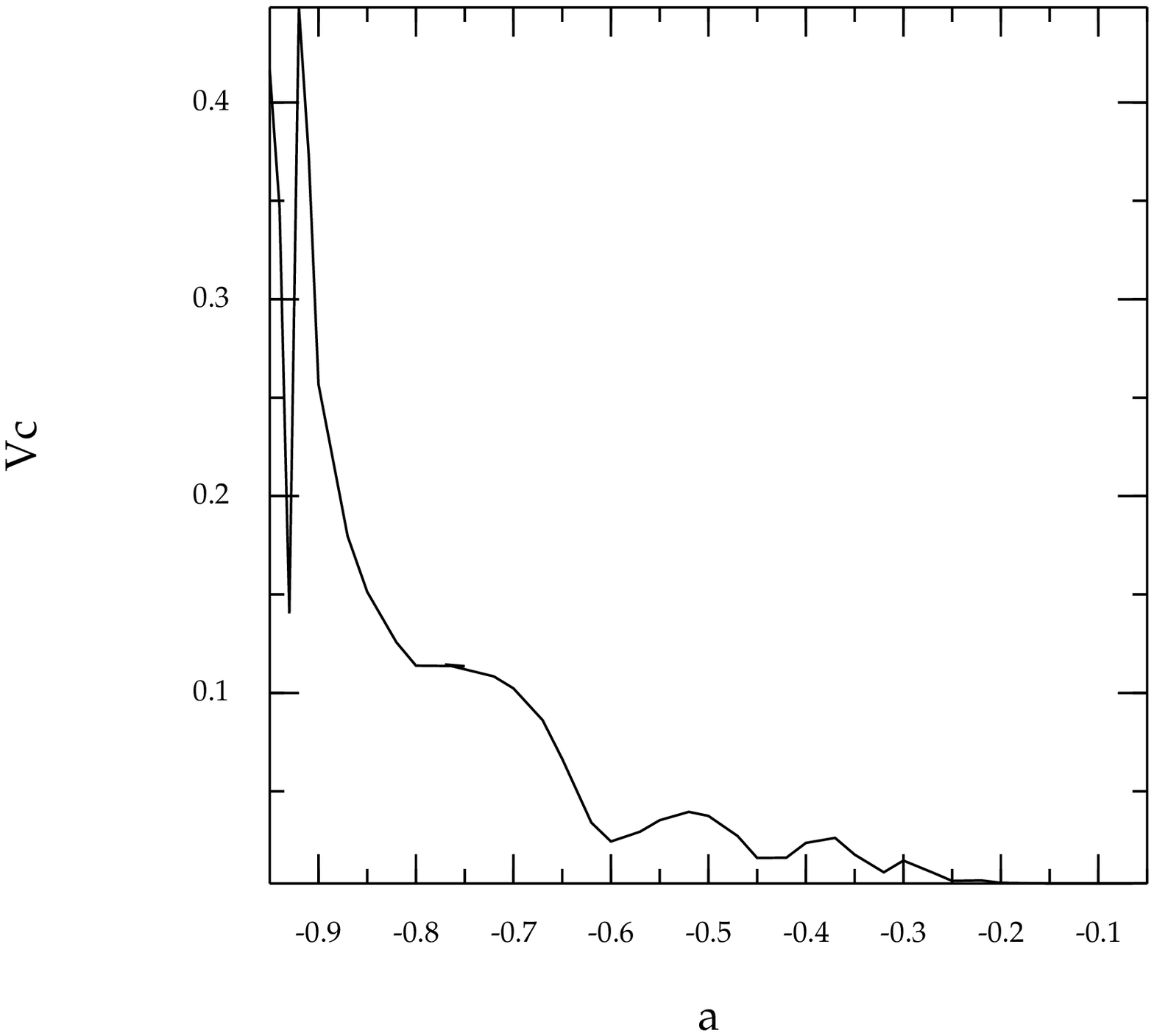}}
 \epsfxsize=8cm \put(8,0){\epsffile{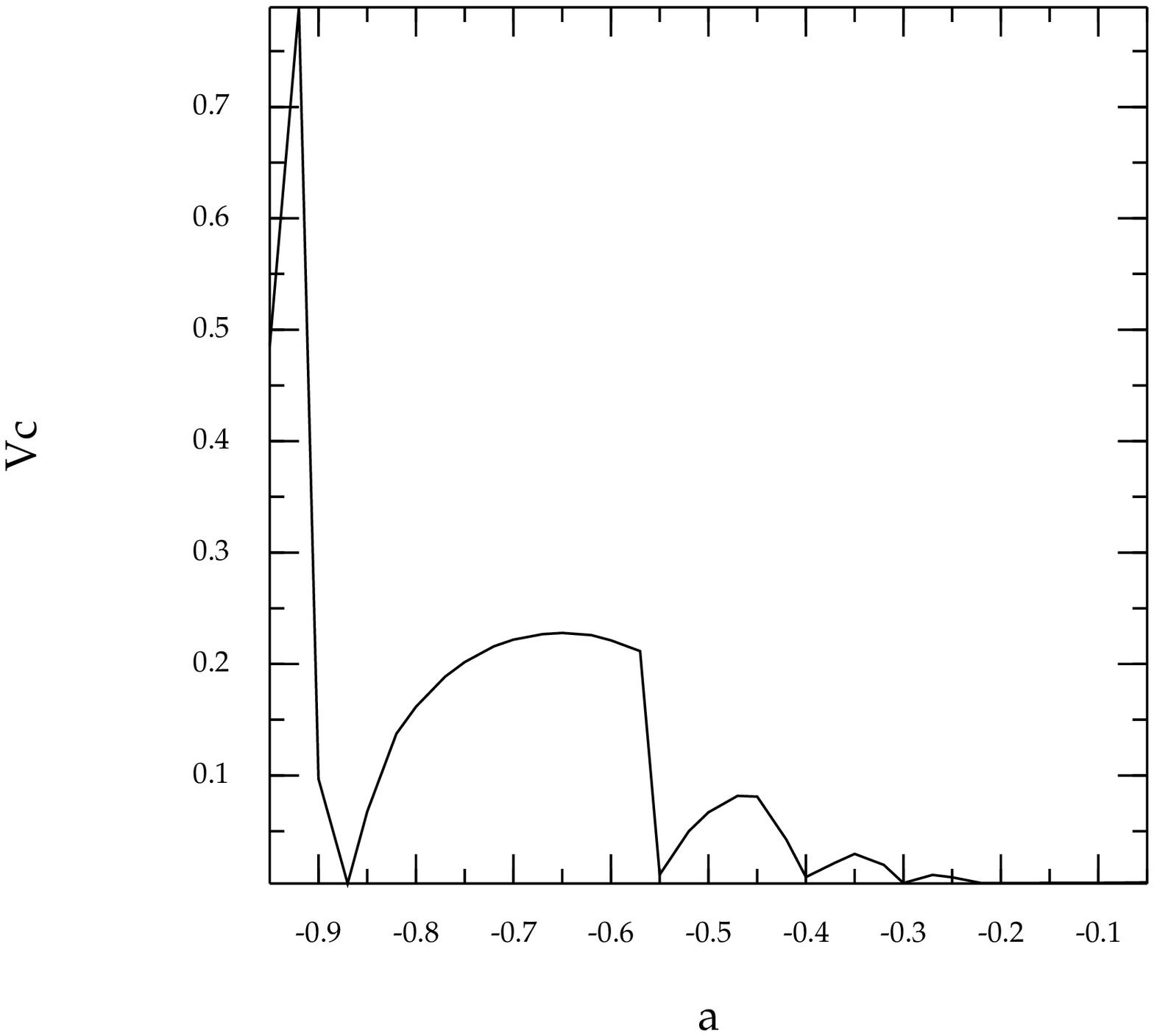}}
\put(4,0){a}
\put(12,0){b}
\end{picture}
\caption{\label{m2vc_L10k} Critical velocity in the 2 mass model 
as a function of the well depth  $a$ for $L=10$ and 
a) $\lambda=0.2$. b) $\lambda=0.5$.
}
\end{figure}

\subsection{Multi-mode Pseudo-Geodesic Model}

The simplest model based on a multi-mode pseudo geodesic approximation would 
involve 3 masses linked 
by two identical springs. For symmetry reasons, the two external masses must 
be identical.
If the frequency ratio of the two vibrational modes that we try to model is of 
the order of 10 percent then the two external masses must be about 10 times 
smaller than the central mass and as a result the system does not absorb 
enough energy to reproduce the scattering of a NBS soliton (the predicted 
critical velocity turns out to be much too small).

\begin{figure}[htbp]
\unitlength1cm \hfil
\begin{picture}(8,8)
 \epsfxsize=8cm \put(0,0){\epsffile{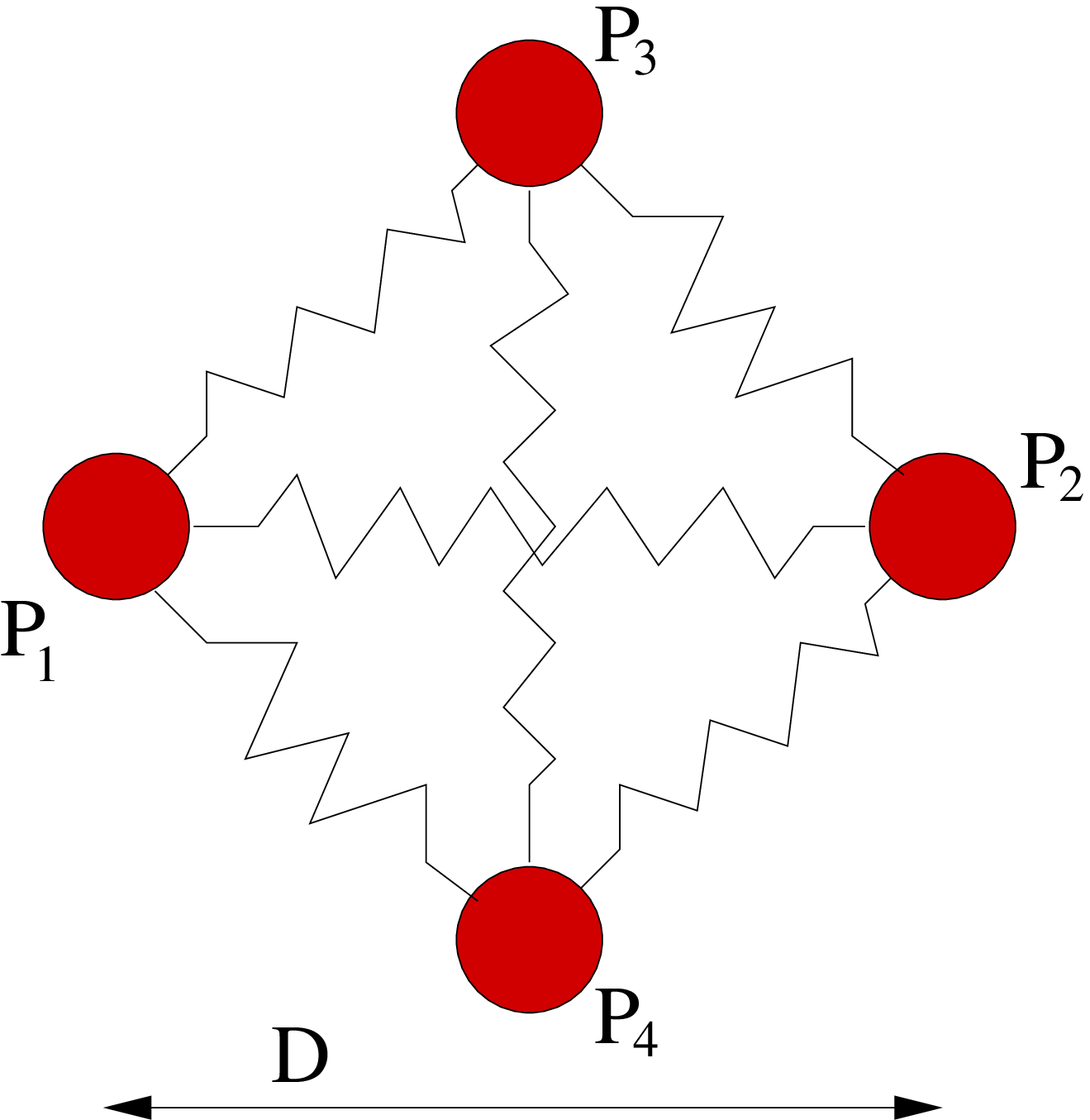}}
\end{picture}
\caption{\label{Model4mass} Configuration of the 4 mass model.
}
\end{figure}

We thus need to construct a model with more degrees of freedom and we have 
chosen 
to use a system made out of 4 identical masses linked by 6 springs in total, 
as pictured in figure
\ref{Model4mass}.

Defining the vector $\vec{Z_i} = (x_i,y_i)$ and the distance 
$R_{ij} = |\vec{Z_i}-\vec{Z_j}|$ we see that 
the Hamiltonian of this system is given by:
\begin{eqnarray}
H &=& \frac{1}{2} (M \dot{\vec{Z_1}}^2 + M \dot{\vec{Z_2}}^2 
               + M \dot{\vec{Z_3}}^2 + M \dot{\vec{Z_4}}^2) 
     +  \frac{k_a}{2} ((R_{12}-D)^2 + (R_{34}-D)^2)
\nonumber\\
 &&  +  \frac{k_b}{2}(R_{13}-\frac{D}{\sqrt{2}})^2 + 
        \frac{k_b}{2}(R_{14}-\frac{D}{\sqrt{2}})^2
     +  \frac{k_b}{2}(R_{23}-\frac{D}{\sqrt{2}})^2 + 
        \frac{k_b}{2}(R_{24}-\frac{D}{\sqrt{2}})^2
\nonumber\\
 &&  +\frac{1}{2}(V(x_1)+V(x_2)+V(x_3)+V(x_4)).
\label{Ham4M}
\end{eqnarray}
Here $k_a$ and $k_b$ are the two parameters which determine the frequencies 
of the 
normal modes of the system and $V(x)$ is the potential (\ref{effPot}).
In addition, $D$ is the equilibrium distance between the 2 pairs of masses and 
is given by 
(\ref{Dx}) where $x_{cm}=(x_1+x_2+x_3+x_4)/4$.
Moreover, $E_0 = 4\, M$ so that the system has the same rest mass as our NBS
soliton.

To see whether this model 
reproduces the results of our simulations 
of the NBS scatterings we must, first of all,  compute the vibration 
frequencies of the system when
the potential $V(x)=0$. This is required to determine the values of $k_a$ and 
$k_b$. To do this, we expand $Z_i$ around its equilibrium position $\bar{Z}_i$:
$x_i = \bar{x}_i+dx_i$ and $y_i = \bar{y}_i+dy_i$. 
To quadratic order, the Hamiltonian (\ref{Ham4M}) becomes
\begin{eqnarray}
H_0 &=&  M (\dot{dx_1}^2 +\dot{dy_1}^2 
         +\dot{dx_2}^2 +\dot{dy_2}^2
         +\dot{dx_3}^2 +\dot{dy_3}^2 
         +\dot{dx_4}^2 +\dot{dy_4}^2
      ) 
\nonumber\\
 &&  + k_a( (dx_1-dx_2)^2 + (dy_3-dy_4)^2) 
\nonumber\\
 &&  + k_b((dx_1-dx_3 + dy_1-dy_3)^2+(dx_1-dx_4 - dy_1+dy_4)^2
\nonumber\\
 &&          +(dx_2-dx_3 - dy_2+dy_3)^2+(dx_2-dx_4 + dy_2-dy_4)^2).
\label{Ham4MLin}
\end{eqnarray}
and so we see that the system posseses 5 vibrational modes and 3 null modes
with the following frequencies and the corresponding eigenmodes: 
\begin{itemize}
\item Breathing mode: $\omega_{br}^2 = 4 k_a/M$; 
  eigen vector: $x_1=-1$, $x_2 = 1$, $y_3=-1$, $y_4=1$.
\item Shape mode:  $\omega_{sh}^2 = (8k_b + 4 k_a)/M$; 
  eigen vector: $x_1=1$, $x_2 = -1$, $y_3=-1$, $y_4=1$.
\item Bend mode 1:  $\omega_{b1}^2 = 8k_b/M$;
  eigen vector: $y_1=-1$, $y_2 = -1$, $y_3=1$, $y_4=1$.
\item Bend mode 2:  $\omega_{b2}^2 = 8k_b/M$; 
  eigen vector: $x_1=-1$, $x_2 = -1$, $x_3 = 1$, $x_4=1$.
\item Squeeze mode:  $\omega_{sq}^2 = 8k_b/M$; 
  eigen vector: $y_1=1$, $y_2=-1$, $x_3 = -1$, $x_4=1$.
\item X translation2:  $\omega_{trx}^2 =0$;
  eigen vector: $x_1=1$, $x_2=1$, $x_3=1$, $x_4 = 1$.
\item Y translation:  $\omega_{try}^2 =0$; 
  eigen vector: $y_1=1$, $y_2=1$, $y_3 = 1$, $y_4 = 1$.
\item Rotation:  $\omega_{rot}^2 =0$;
  eigen vector: $y_1=-1$, $y_2=1$, $x_3=-1$, $x_4 = 1$.
\end{itemize}

To model the NBS solitons, we fix the parameters $k_b$ and $k_a$ by 
fitting $\omega_{br}$ and $\omega_{sh}$ to the values of the 
corresponding frequencies of a NBS soliton: 
$\omega_{br} = (0.3+0.63(1-exp(-1.58 \lambda)))^{1/2}$. Choosing
$k_b = K_{ab} k_a$ we then take
$k_a = \frac{M}{4} (0.3+0.63(1-exp(-1.58 \lambda)))^{1/2}$ and
$\omega_{sh} = K_{\omega} \omega_{br}$. Using the expressions for the normal modes
$\omega_{sh}$ and $\omega_{br}$ we have $ K_{ab} = (k_\omega^2-1)/2$. 

Moreover, the effective value of $\lambda$ varies when the soliton falls into
the whole. To take this effect into account, we introduce the
profile 
\begin{equation}
\lambda_{eff} = \lambda (1 -\frac{1}{2}((1+\alpha)^{1/2} -1) P(S_{\lambda}; x_cm))
\label{leff}
\end{equation}
which we then use in the expression for $k_a$:
\begin{equation}
k_a = \frac{M}{4} (0.3+0.63(1-exp(-1.58 \lambda_{eff})))^{1/2},
\label{kaprof}
\end{equation}
making it position dependant. 

If we define
\begin{equation}
Q =  \frac{1}{8} ((R_{12}-D)^2 + (R_{34}-D)^2 + K_{ab}(
        (R_{13}-\frac{D}{\sqrt{2}})^2 + 
        (R_{14}-\frac{D}{\sqrt{2}})^2
     +  (R_{23}-\frac{D}{\sqrt{2}})^2 + 
        (R_{24}-\frac{D}{\sqrt{2}})^2)) \frac{d k_a }{ d x_{cm}}
\end{equation}
then the equations of motion for the degrees of freedom of our Hamiltonian 
(\ref{Ham4M}) are given by

\begin{eqnarray}
M \ddot{x_1} &=& 
      -k_a (R_{12}-D){(x_1-x_2)\over R_{12}}
      -k_a(R_{12}+R_{34}-2 D)\frac{1}{8}\,d\,G(s_{D};x_{cm})
\nonumber\\
    &&-k_b (R_{13}-\frac{D}{\sqrt{2}}) {(x_1-x_3)\over R_{13}}
      -k_b (R_{14}-\frac{D}{\sqrt{2}}) {(x_1-x_4)\over R_{14}}
      \nonumber\\
    &&-k_b(R_{13}+R_{14}+R_{23}+R_{24}-4 \frac{D}{\sqrt{2}})
            \frac{1}{8}\,d\,G(s_{D};x_{cm})
      - Q - \frac{dE}{4} G(s;x_1)
\nonumber\\
M \ddot{y_1} &=& -k_a (R_{12}-D){(y_1-y_2)\over R_{12}}
             - k_b (R_{13}-\frac{D}{\sqrt{2}}){(y_1-y_3)\over R_{13}}
             - k_b (R_{14}-\frac{D}{\sqrt{2}}){(y_1-y_4)\over R_{14}}\nonumber\\
\nonumber\\
M \ddot{x_2} &=& 
      -k_a (R_{12}-D){(x_2-x_1)\over R_{12}}
      -k_a(R_{12}+R_{34}-2 D)\frac{1}{8}\,d\,G(s_{D};x_{cm})
\nonumber\\
    &&- k_b (R_{13}-\frac{D}{\sqrt{2}}) {(x_2-x_3)\over R_{23}}
      - k_b (R_{14}-\frac{D}{\sqrt{2}}) {(x_2-x_4)\over R_{24}}
\nonumber\\
    &&- k_b(R_{13}+R_{14}+R_{23}+R_{24}-4 \frac{D}{\sqrt{2}})
            \frac{1}{8}\,d\,G(s_{D};x_{cm})
      - Q - \frac{dE}{4} G(s;x_2)
\nonumber\\
M \ddot{y_2} &=& -k_a (R_{12}-D){(y_2-y_1)\over R_{12}}
               -k_b (R_{13}-\frac{D}{\sqrt{2}}){(y_2-y_3)\over R_{13}}
               -k_b (R_{14}-\frac{D}{\sqrt{2}}){(y_2-y_4)\over R_{14}}\nonumber\\
\nonumber\\
M \ddot{x_3} &=& 
     -k_a (R_{34}-D)({(x_3-x_4)\over R_{34}}
     -k_a (R_{12}+R_{34}-2 D)\frac{1}{8}\,d\,G(s_{D};x_{cm})
\nonumber\\
    &&- k_b (R_{31}-\frac{D}{\sqrt{2}}) {(x_3-x_1)\over R_{31}}
      - k_b (R_{32}-\frac{D}{\sqrt{2}}) {(x_3-x_2)\over R_{32}}
\nonumber\\
    &&- k_b(R_{13}+R_{14}+R_{23}+R_{24}-4 \frac{D}{\sqrt{2}})
            \frac{1}{8}\,d\,G(s_{D};x_{cm})
      - Q - \frac{dE}{4} G(s;x_3)
\nonumber\\
M \ddot{y_3} &=&- k_a (R_{34}-D){(y_3-y_4)\over R_{34}}
              - k_b (R_{31}-\frac{D}{\sqrt{2}}) {(y_3-y_1)\over R_{31}}
              - k_b (R_{32}-\frac{D}{\sqrt{2}}) {(y_3-y_2)\over R_{32}}
\nonumber\\
M \ddot{x_4} &=& 
     -k_a (R_{34}-D) {(x_4-x_3)\over R_{34}}
     -k_a (R_{12}+R_{34}-2 D)\frac{1}{8}\,d\,G(s_{D};x_{cm})
\nonumber\\
    && - k_b (R_{41}-\frac{D}{\sqrt{2}}) {(x_4-x_1)\over R_{41}}
       - k_b (R_{42}-\frac{D}{\sqrt{2}}) {(x_4-x_2)\over R_{42}}
\label{4meq}\\
    &&- k_b(R_{13}+R_{14}+R_{23}+R_{24}-4 \frac{D}{\sqrt{2}})
            \frac{1}{8}\,d\,G(s_{D};x_{cm})
       - Q - \frac{dE}{4} G(s;x_4)
\nonumber\\
M \ddot{y_4} &=& - k_a (R_{34}-D){(y_4-y_3)\over R_{34}}
              - k_b (R_{41}-\frac{D}{\sqrt{2}}){(y_4-y_1)\over R_{41}}
              - k_b (R_{42}-\frac{D}{\sqrt{2}}){(y_4-y_2)\over R_{42}}.
\nonumber
\end{eqnarray}

We could also add various friction terms but, to keep the model simple, we 
have just added twi single friction terms proportional to 
$\dot{x_2}-\dot{x_1}$ and $\dot{y_3}-\dot{y_4}$. 
We thus assume that only the diagonal springs can 
radiate. In the new baby Skyrme model, the lowest vibrational mode
of a single soliton is coupled to the radiation when 
$\lambda < \lambda_{c_{min}} = 0.27$. As $\lambda = w_c^2$, for higher vibration 
frequency the critical value of $\lambda$ is given by
$\lambda_c = K_\omega^2 \lambda_{c_{min}}$.
We have thus added the following friction force when $\lambda_{eff} < \lambda_c$:
\begin{eqnarray}
Fr_{x_1} &=&  -K_{fr} (\lambda_c - \lambda_{eff})(\dot{x_1}-\dot{x_2})
\nonumber\\ 
Fr_{x_2} &=&  -K_{fr} (\lambda_c - \lambda_{eff})(\dot{x_2}-\dot{x_1})
\nonumber\\ 
Fr_{y_3} &=&  -K_{fr} (\lambda_c - \lambda_{eff})(\dot{y_3}-\dot{y_4})
\nonumber\\ 
Fr_{y_4} &=&  -K_{fr} (\lambda_c - \lambda_{eff})(\dot{y_4}-\dot{y_3}),
\end{eqnarray}
where $\lambda_{eff}$ is given by (\ref{keff}) and  $K_{fr}$ is a friction 
coefficient.
We have also tried to add some friction along the lateral strings, but 
we have found 
that this did not model well the dissipation of energy by the soliton
so from now on we set such friction to zero. 

\begin{figure}[htbp]
\unitlength1cm \hfil
\begin{picture}(16,8)
 \epsfxsize=8cm \put(0,0){\epsffile{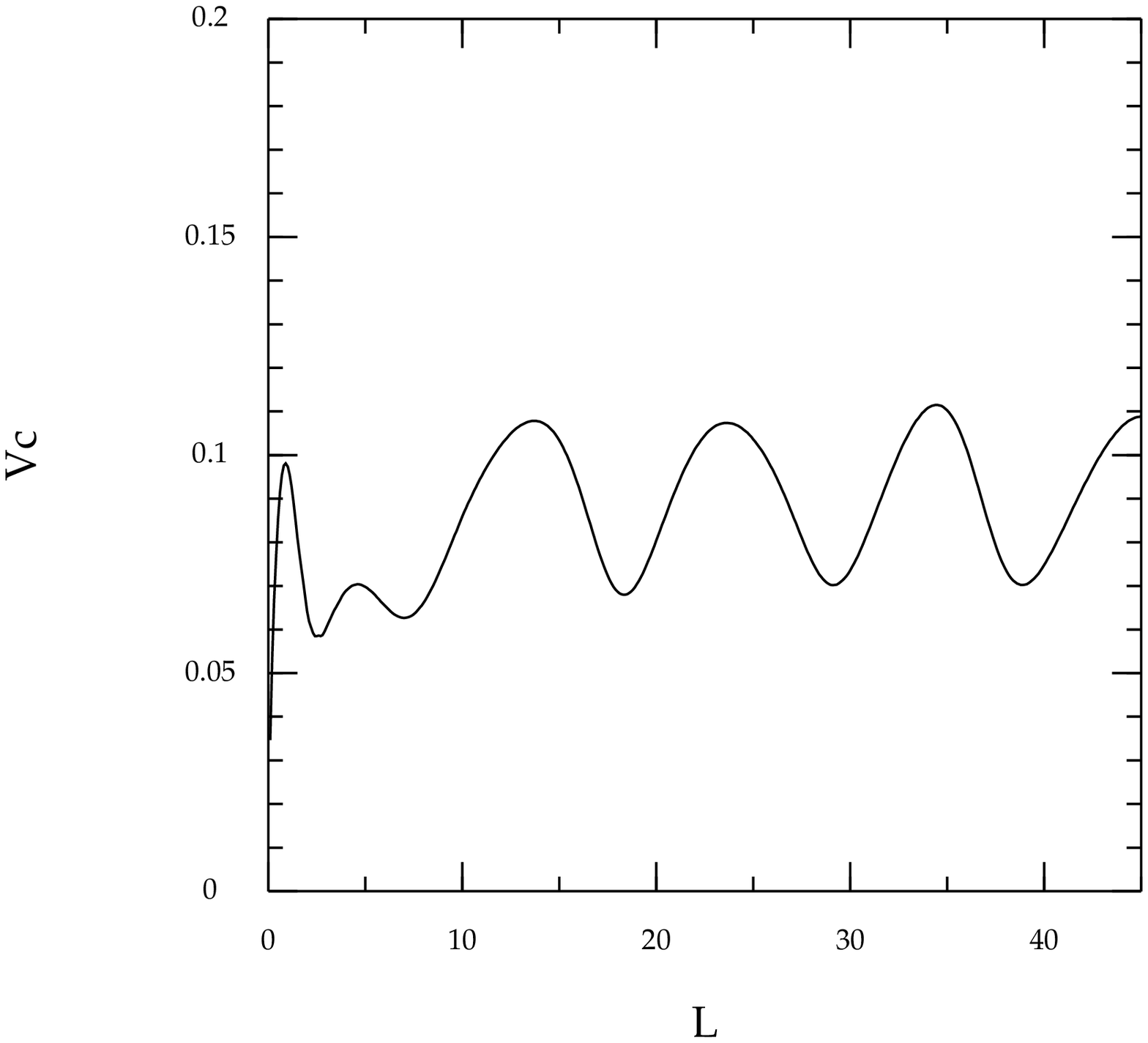}}
 \epsfxsize=8cm \put(8,0){\epsffile{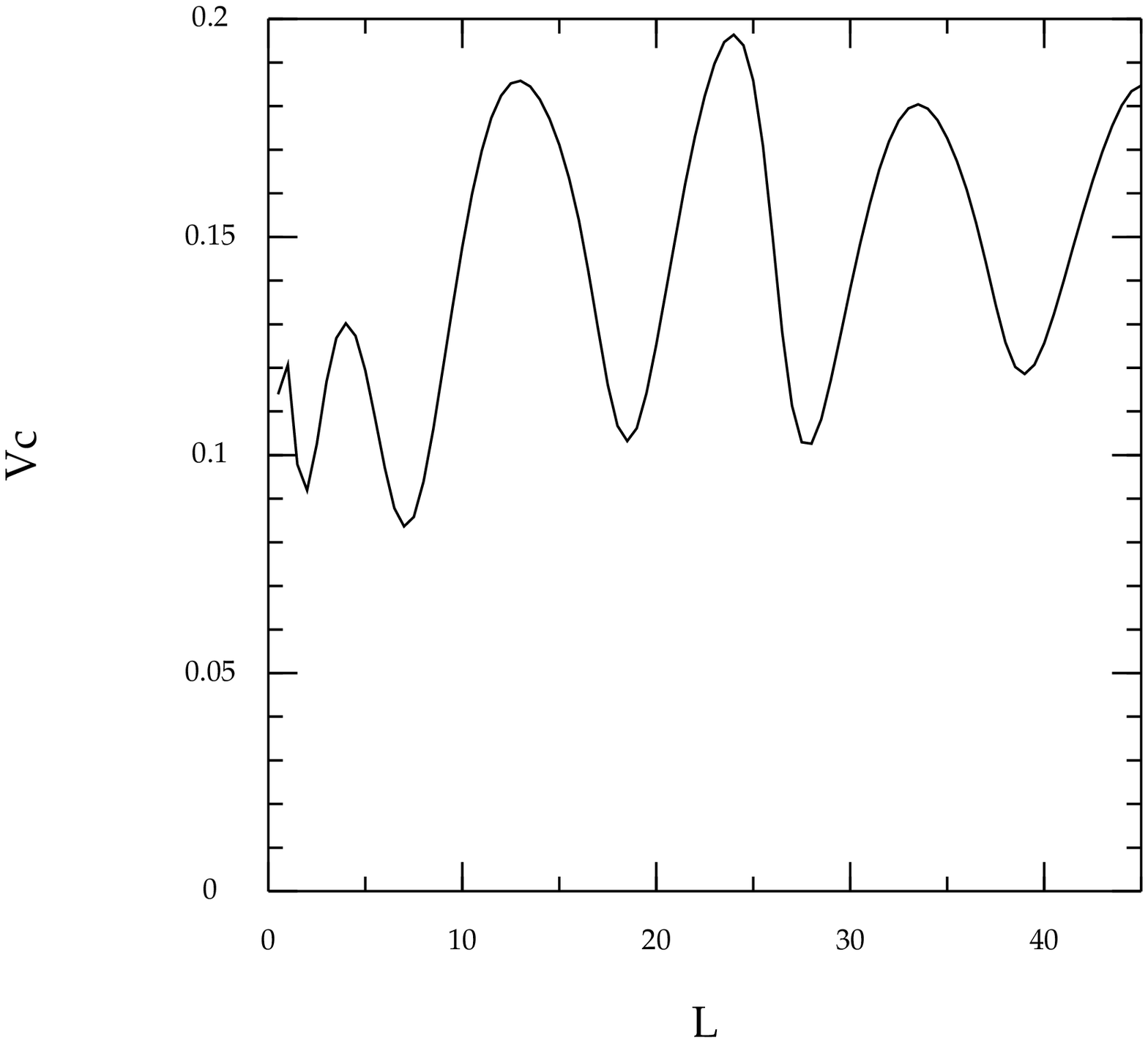}}
\put(4,0){a}
\put(12,0){b}
\end{picture}
\caption{\label{m4vc_a-0.8k} Critical velocity in the 4 mass model 
as a function of the well width $L$ for $a=-0.8$ and 
a) $\lambda=0.2$, $s=s_D=s_k=0.4, K_{fr}=0.02, K_\omega=1.45$ 
b) $\lambda=0.5$, $s=s_D=s_k=0.47, K_{fr}=0.025, K_\omega=1.63$.
}
\end{figure}

Solving numerically the equations of motion (\ref{4meq}) we found that the
scale factor $s$, $s_D$ and $s_\lambda$ can be set to the same value, 
as one would expect.
Trying to reproduce the data shown in Fig. \ref{vc_a-0.8k} from the full 
2 dimensional equations,
we have found that the best values of the model parameters are given by
\begin{itemize}
\item{$\lambda=0.2$}: $s = s_D = s_\lambda = 0.4$, $K_\omega = 1.45$ 
and $K_{fr}=0.025$.
\item{$\lambda=0.5$}: $s = s_D = s_\lambda = 0.47$, $K_\omega = 1.63$  
and $K_{fr}=0.02$.
\end{itemize}
Figure \ref{m4vc_a-0.8k} shows the dependence of the critical velocity 
as a function of the well width $L$ for the 4 mass model.

Like the 2 mass model, the 4 mass model predicts that  the value of the 
critical 
velocity oscillates as the well width $L$ increases. 
The period of oscillations are nevertheless larger than in the 2 mass model
and is controlled in our model mainly by the parameter $K_\omega$. 
The amplitude of 
oscillations and the minimum of the oscillations are on the other hand 
determined mostly by the parameters $s$ and $K_{fr}$. 
When  $K_{fr}$ is small, the critical velocity is very small
but the oscillations are then very irregular.
 
The 4 mass model reproduces the correct range of quantities describing the 
oscillation for both $\lambda=0.2$ and $\lambda=0.5$. Though the curves in
Fig \ref{m4vc_a-0.8k} and Fig \ref{vc_a-0.8k} do not march perfectly, their
general features are similar. Even the small hump just below $L=5$,
is more or less reproduced.

\begin{figure}[htbp]
\unitlength1cm \hfil
\begin{picture}(16,8)
 \epsfxsize=8cm \put(0,0){\epsffile{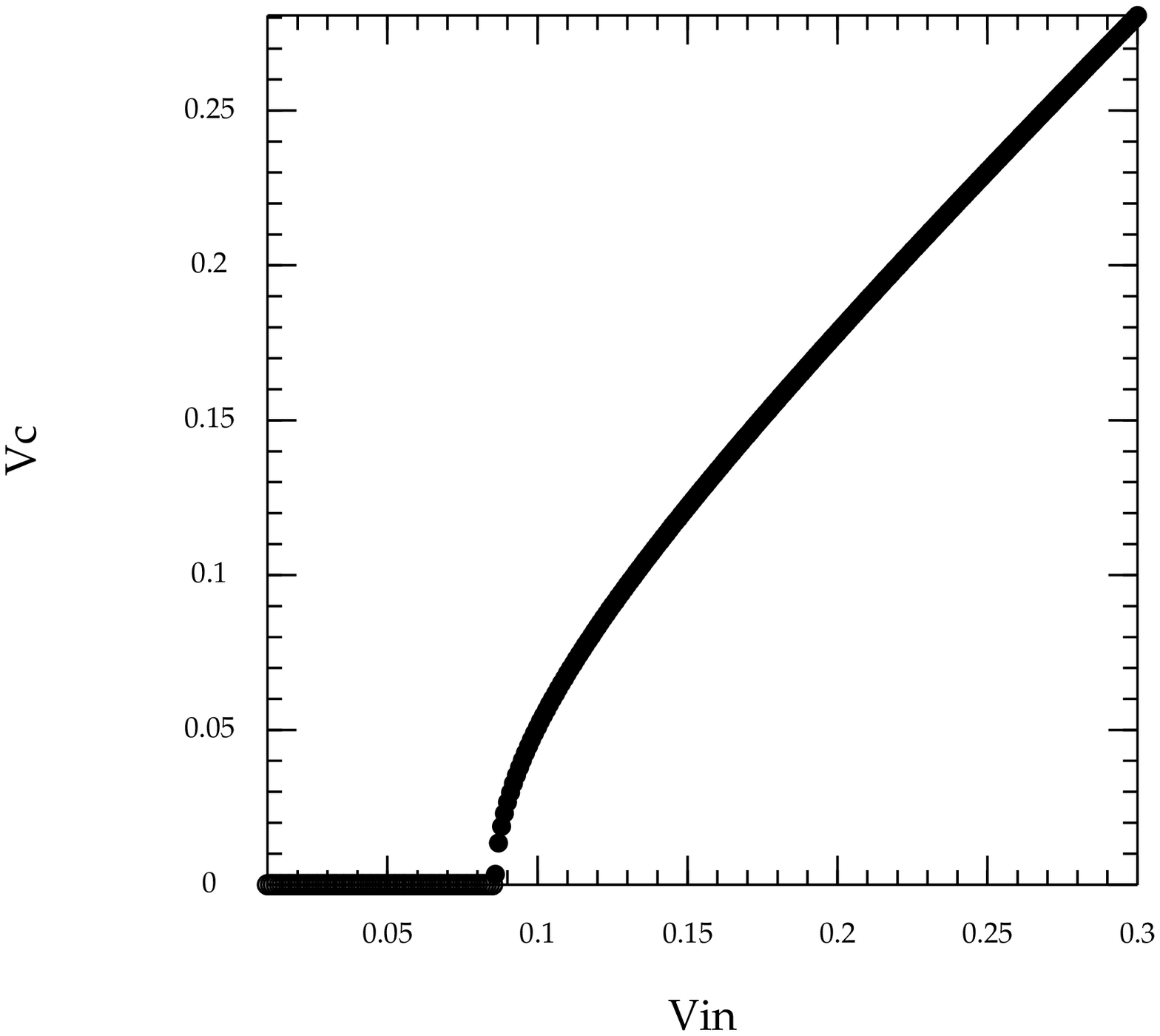}}
 \epsfxsize=8cm \put(8,0){\epsffile{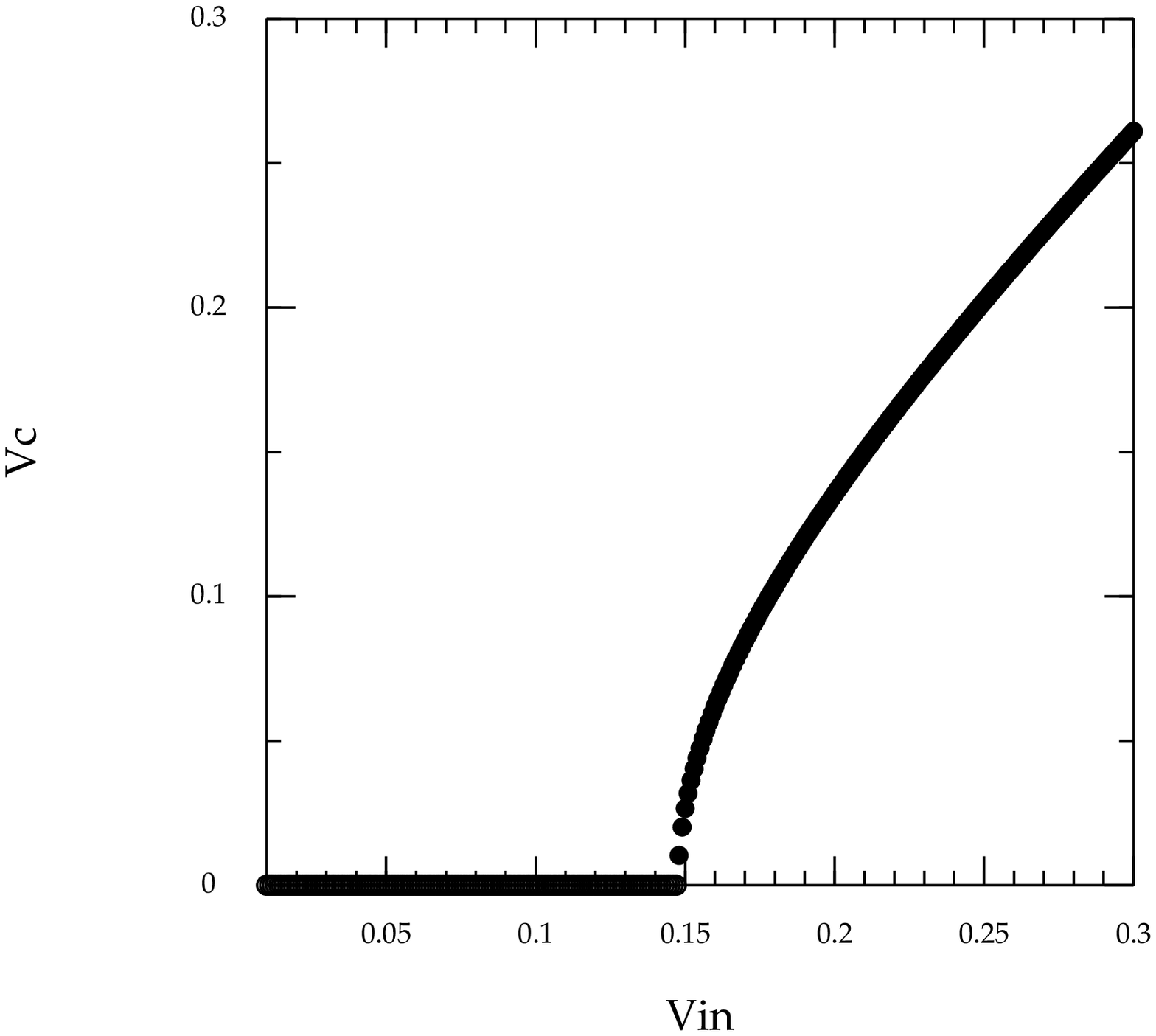}}
\put(4,0){a}
\put(12,0){b}
\end{picture}
\caption{\label{m4vout_L10k} Speed of a soliton in the 4 mass model after 
scattering on a well of width $L=10$ and depth $a=-0.8$ for 
a) $\lambda=0.2$. b) $\lambda=0.5$. [To be compared with Fig. \ref{vout_L10k}]
}
\end{figure}

\begin{figure}[htbp]
\unitlength1cm \hfil
\begin{picture}(16,8)
 \epsfxsize=8cm \put(0,0){\epsffile{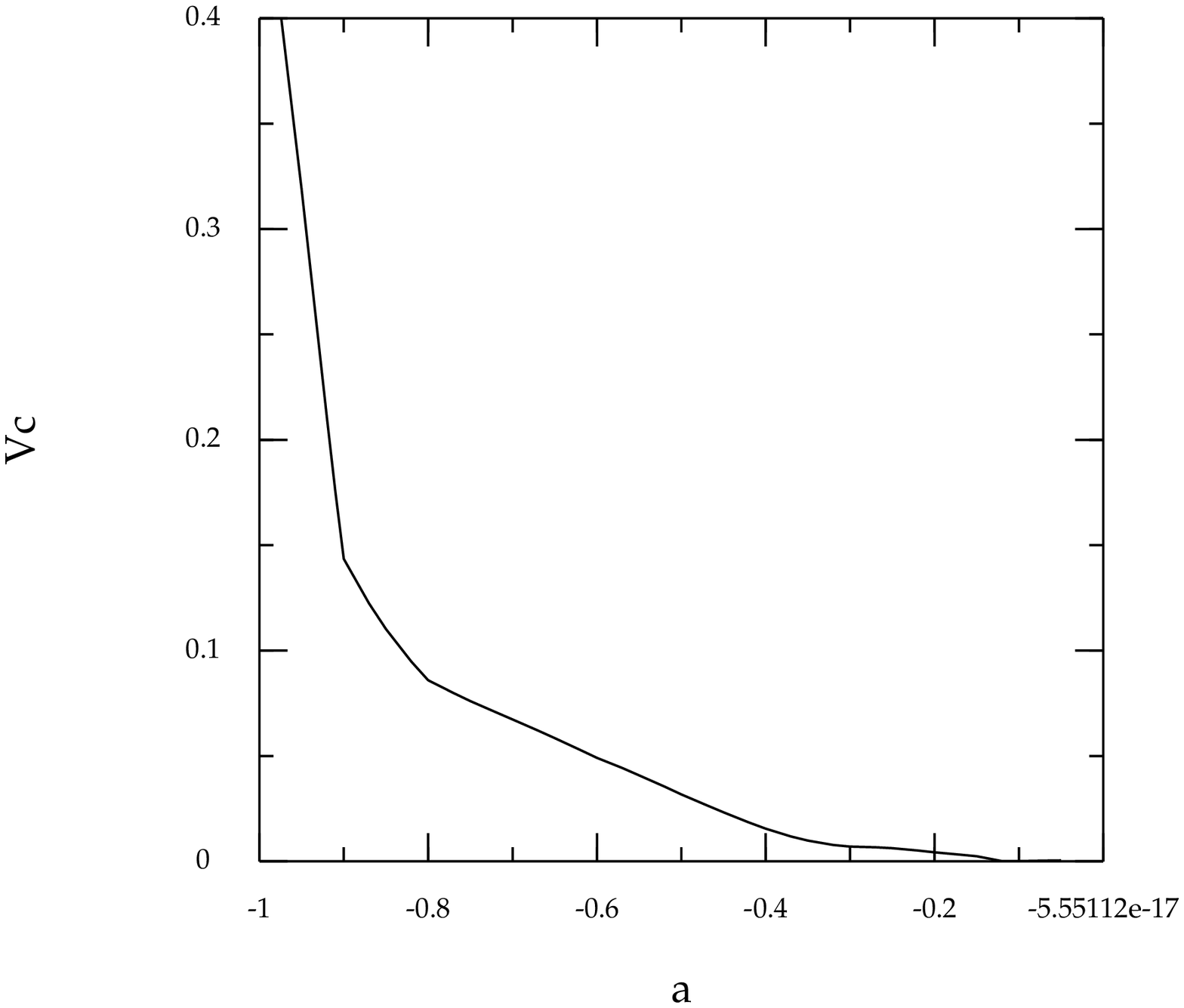}}
 \epsfxsize=8cm \put(8,0){\epsffile{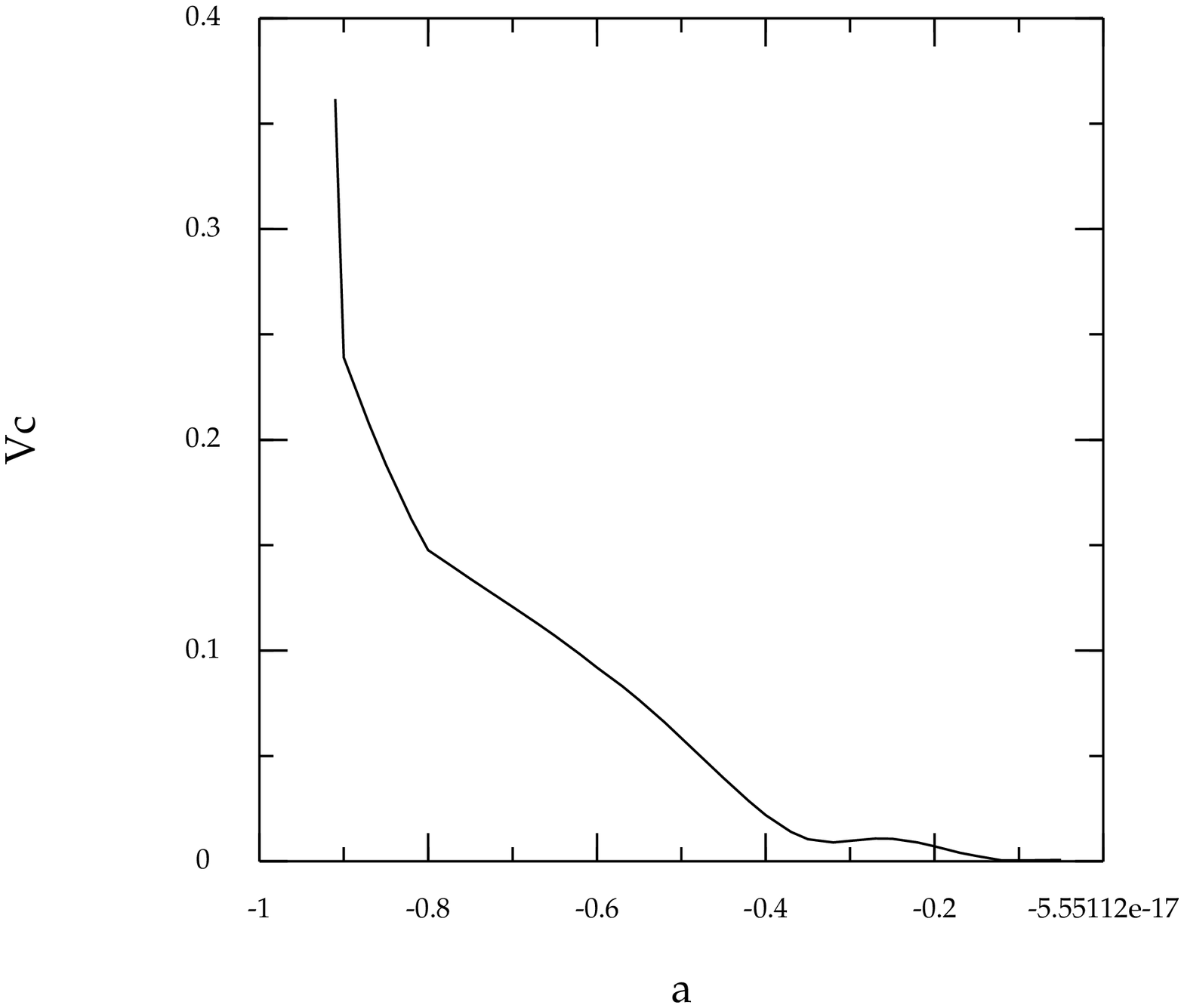}}
\put(4,0){a}
\put(12,0){b}
\end{picture}
\caption{\label{m4vc_L10k} Critical velocity in the 4 mass model 
as a function of the well depth  $a$ for $L=10$ and 
a) $\lambda=0.2$. b) $\lambda=0.5$. [To be compared with Fig. \ref{vc_L10k}]
}
\end{figure}

Having set the parameters of the 4 mass model, we then obtained the 
figures \ref{m4vout_L10k} and \ref{m4vc_L10k} 
which should be compared with figures \ref{vout_L10k}, and \ref{vc_L10k}
respectively.

We notice in particular in Fig. \ref{m4vout_L10k} that the relation between 
the ingoing and outgoing velocity are very similar, even if the predicted 
critical velocity is slightly smaller than the real one. 
When $\lambda=0.2$ the outgoing velocity has a small dip around $v_{in}=0.15$,
Fig \ref{vout_L10k} a, a feature that our 4 mass model does not reproduce. 
We believe that this is caused by a subtle interference between the various 
oscillation modes of the model, which our 4 mass model cannot reproduce.

Figure \ref{m4vc_L10k} presents the dependence of the critical velocity as a 
function of the depth of the well $a$. The model works surprisingly 
well when $\lambda=0.5$ reproducing the gross features of Figure 
\ref{vc_L10k} except
when $a$ is close to $-1$. In this region, the soliton is very broad when it is
inside the well and so its overlap with the edges of the well is not properly 
taken into account by our pseudo-geodesic approximation.

When $\lambda=0.2$ the dependence of the critical speed on the parameter $a$
predicted by the pseudo-geodesic approximation is very crude. In particular, 
we do not reproduce the plateau where $V_c$ is constant for small values of 
$|a|$. We attribute this to the important role of the dispersion of energy 
through radiation that our pseudo-geodesic approximation takes into account 
only very crudely.

The fact that the four mass pseudo-geodesic approximation reproduces the 
properties of the NBS soliton much better than the 2 mass model shows very 
clearly that it is the interference between several vibrational modes of the 
soliton that determines the scattering properties of the soliton on the 
well. It is the relative phases between these vibrational
modes that determine the variation of the critical velocity when the 
parameters of the model are changed.

The pseudo-geodesic approximation reproduces well the observation that when 
the soliton 
falls into the well it gets excited.
A fraction of its kinetic energy is transferred to oscillation modes. To escape
from the well, the soliton needs to be in a correct phase, meaning that all 
its oscillations must be in a correct phase. 
The fact that the parameter $K_\omega$ is larger than $1.1$ shows that it is
not just the two lowest vibrational modes that matter, but that the higher 
pseudo vibrational modes, which radiate, play a key role as well.

\section{Conclusion}
We have shown that the scattering of the NBS soliton on a well has several
interesting properties. At large speeds, the soliton crosses the well
nearly elastically, while at small speeds it is trapped by the well.
For a very small range of values of the parameters, the soliton can sometimes 
bounce a few times inside the well before escaping from it.

The major factor in the interaction between the well and the soliton comes from
the fact that the size of the soliton at rest inside the well is different 
from its size outside the well. Moreover, the energy of the soliton is smaller 
inside the well. As the soliton falls into the well, it alters its 
shape and as a result, its vibrational modes become exited. These oscillations
absorb some energy from the soliton, making it impossible for the soliton to 
escape from the well unless it has enough kinetic energy to make up for the
amount of energy stored in the vibrational modes.

Near the critical velocity, the behaviour of the soliton is dictated by the
 various vibrational modes that get excited during the scattering process. 
For the soliton to escape from the well, these vibrations must be in the right 
phase when the soliton tries to climb out of the well. 
This is illustrated by the fact that the critical velocity is a 
periodic function of $L$ (Fig \ref{vc_a-0.8k}).

To verify this point, we have presented 2 models based  on a pseudo-geodesic 
approximation to the full process. Both models approximately
describe the evolution of the soliton in the well and describe 
some of its vibrational modes. 
We have found that to describe the scattering of the soliton on the well 
reasonably 
accurately the effective models must take into account at least 2 vibrational 
modes of the soliton, one of which corresponds to a vibrational mode that 
radiates.  

We have thus shown that the scattering of solitons on a well has interesting 
properties that need to be investigated further. Many other models  
similar to the NBS model can also have wells and it would be interesting to 
find out if they exhibit similar phenomena to the ones that we have described 
in this paper
and if the vibrational modes of the solitons always play such a crucial role.

\section{Acknowledgement}

This investigation is a natural follow up of the work \cite{PZB} performed
in collaboration with Joachim Brand. We would like to thank him 
for this collaboration.
The work reported in this paper was, in part, supported by a 
PPARC grant  (PPA/G/S/2003/00161).

\end{document}